\documentclass[conference,compsoc]{IEEEtran}
\usepackage{booktabs,tabularx,makecell}

\usepackage{multirow}
\usepackage{makecell} 

\usepackage{listings}

\usepackage{xspace} 
\newcommand{\sys}{\textsc{PCodeTrans}\xspace} 

\usepackage{xcolor}
\usepackage[skins,breakable,most]{tcolorbox}
\newtcolorbox{promptbox}[1][]{
  colback=gray!5!white,
  colframe=black!75!white,
  title=\textbf{System Prompt},
  fonttitle=\bfseries,
  boxrule=0.5mm,
  breakable,   
  #1
}

\usepackage{inconsolata}

\usepackage{enumitem}
\setlist[itemize]{leftmargin=1em}

\ifCLASSOPTIONcompsoc
  \usepackage[nocompress]{cite}
\else
  \usepackage{cite}
\fi
\ifCLASSINFOpdf
  \usepackage{graphicx}
\else
\fi
\usepackage{url}

\begin{document}
%
\title{PCodeTrans: Translate Decompiled Pseudocode to Compilable and Executable Equivalent}

    
    

\author{
    \IEEEauthorblockN{Yuxin Cui\textsuperscript{1}, Zeyu Gao\textsuperscript{1}, Shuxian He\textsuperscript{2}, Siliang Qin\textsuperscript{3}, Chao Zhang\textsuperscript{1,*}}
    \IEEEauthorblockA{\textsuperscript{1}Tsinghua University, \textsuperscript{2}Wuhan University, \textsuperscript{3}Institute of Information Engineering, CAS}
    \IEEEauthorblockA{
        \texttt{\{yx-cui24, gaozy22\}@mails.tsinghua.edu.cn, chaoz@tsinghua.edu.cn} \\
        \texttt{heshuxian@whu.edu.cn, qinsiliang@iie.ac.cn}
    }
    \IEEEauthorblockA{\textsuperscript{*}Corresponding Author}
}

\maketitle

\begin{abstract}
Decompilation is foundational to binary analysis, yet conventional tools prioritize human readability over strict recompilability and verifiable runtime correctness. 
While recent LLM-based approaches attempt to refine decompiled pseudocode, they typically either optimize solely for readability or rely on static analysis for evaluation. This makes them prone to ``semantic hallucinations'' that compromise accuracy and fail to resolve actual runtime failures. For critical tasks like software modernization and vulnerability remediation, recovered code must not only compile but replicate the original binary's behavior. 
We present \sys, a feedback-driven framework that bridges the gap between decompilation, recompilation, and rigorous function-level dynamic validation. After extracting a minimal yet coherent context to guarantee recompilability, \sys employs an in-situ substitutable engine to hot-swap the compiled function directly into the unmodified binary, natively preserving its authentic execution context and global dependencies.
Guided by fine-grained differential tracing, \sys generates precise runtime feedback to iteratively guide an LLM in repairing semantic discrepancies. 
Evaluated on \textit{Coreutils} and \textit{Binutils}, \sys achieves unprecedented recovery performance when rectifying raw Hex-Rays outputs, attaining 100\% function-level compilability on unstripped binaries alongside 99.55\% and 99.89\% test-validated behavioral consistency, respectively. In doing so, it resolves 76.56\% and 79.74\% of logic errors exposed by official test suites. Exhibiting exceptional resilience, \sys maintains over 96\% behavioral consistency even on fully stripped binaries. By significantly outperforming all existing baselines, \sys paves a practical path to reliably translate decompiled pseudocode into compilable and executable equivalents.
\end{abstract}


%
\IEEEpeerreviewmaketitle

\section{Introduction}

As modern software increasingly ships without accessible source code, critical workflows in security auditing~\cite{mantovani2022CSCandBVD,han2023queryx,burk2022decomperson} and software modernization~\cite{Reiter2025autoMitiVuln,khadka2015softmodcasestudy} often start from binaries. Decompilation offers a practical path to recover high-level, C-like pseudocode with limited readability, helping analysts reason about program logic rather than the minutiae of machine instructions. However, for downstream tasks such as vulnerability discovery with source-based analyzers (e.g., CodeQL~\cite{CodeQL} or Joern~\cite{Joern2014}) and large-scale software modernization, this level of quality remains insufficient. The recovered code should be able to be recompiled by standard toolchains and behave equivalently to the original binary when executed~\cite{Wong2025decllm,Zou2025d-lift,Chen2025recopilot}.

Rule-based decompilers~\cite{idapro,ghidra,retdec} inevitably rely on handcrafted recovery rules~\cite{Zou2024DHelix} and heuristic guesses to reconstruct source-level information that has been erased or distorted by compiler optimizations. Their outputs frequently trigger syntax-related compiler errors under \texttt{gcc} or \texttt{g++}~\cite{Wang2023sem2vec,Ji2023benchllmcodegen}. Even when compilation succeeds, the behavior of the resulting code can diverge from the binary at runtime~\cite{david2018firmup}. For example, Hex-Rays~\cite{idapro} often suffers from misrecovered or incompatible types, incorrect function prototypes (e.g., \texttt{varargs}, struct returns), uninitialized variables, and non-standard constructs such as tool-specific intrinsics and opaque jump-table encodings that mirror disassembly artifacts rather than compilable, source-level semantics~\cite{liu2023decompx86,Chen2022dirty}. Other off-the-shelf rule-based decompilers, including Ghidra~\cite{ghidra} and RetDec~\cite{retdec}, exhibit similar problems and frequently produce outputs that are not readily compilable and that diverge semantically from the original binary~\cite{Li2024splitmerge,Zhou2024Plankton}.

Recent LLM-based refinement approaches~\cite{Hu2024DeGPT,Tan2024llm4decompile,Lacomis2019dire,Nitin2021direct,Tan2025sk2decompile} mainly target readability. They make pseudocode look more like human-written sources by polishing decompiled functions~\cite{Hu2024DeGPT,Tan2024llm4decompile}, recovering names, variables, and types~\cite{Lacomis2019dire,Nitin2021direct}, or combining both directions~\cite{Tan2025sk2decompile}. However, these readability gains often come at the cost of reduced usability. Some pipelines attempt to “clean up’’ decompiled code by iterating on compiler diagnostics and coarse runtime signals~\cite{Wong2025decllm}, while others introduce training signals such as a score that combines syntactic correctness with a notion of semantic correctness derived from symbolic execution and SMT solving~\cite{Zou2025d-lift}. 
However, these approaches either struggle to scale to complex production binaries, or they suffer from path explosion and depend heavily on the precise modeling of complex execution environments—including heap state, global variables, and external libraries. Consequently, across these efforts, the verification of semantic correctness remains limited~\cite{pei2021stateformer}.

To address these limitations, we design \sys around three core challenges and their corresponding solutions. 
First, context-unaware function-level recompilation is fragile. 
Raw decompiler output naturally lacks critical dependencies, and independently relying on LLMs to infer them across isolated functions often fails~\cite{Tan2024llm4decompile}.
\sys overcomes this by reconstructing a unified, globally shared compilation context before any local repair. By anchoring each function's recompilation to this consistent foundation, \sys ensures global compatibility while leveraging compiler diagnostics to guide the LLM in an iterative process to resolve local failures.
Second, verifying behavioral equivalence for compiled functions remains challenging: manual test construction does not scale, while symbolic and SMT-based methods face path explosion and complex environment modeling\cite{Zou2025d-lift}. 
Instead, \sys introduces \emph{in-situ} substitutable execution, which hot-swaps the single function into the unmodified binary. This allows the program's own test executions to serve as a practical equivalence oracle.
Third, knowing a function fails a test is insufficient for repair, as I/O mismatches or symbolic counterexamples are too coarse to localize root causes\cite{Zhou2026fidelitygpt}. To enable repair, \sys employs runtime feedback–guided repair, combining sanitizer diagnostics with Breakpoint-matched Differential Tracing. This technique compares values and control flow at aligned breakpoints between original and substituted executions, generating fine-grained evidence that guides the LLM to pinpoint and correct errors.

We implement \sys on Linux (x86\_64), building on Hex-Rays for decompilation and state-of-the-art LLMs for refinement. For evaluation, we construct datasets from two established GNU C suites, \texttt{coreutils-9.5}~\cite{coreutils} and \texttt{binutils-2.45}~\cite{binutils}. To comprehensively evaluate real-world applicability, we assess both standard (unstripped) and stripped versions of these binaries, measuring correctness via each suite's official regression harness. \sys markedly outperforms all existing approaches. On the unstripped datasets, it achieves 100\% compilability and over 99.5\% behavioral equivalence, automatically resolving over 79\% of latent semantic errors exposed in the raw Hex-Rays outputs. Even under challenging stripped settings, \sys maintains over 96\% behavioral equivalence while repairing over 58\% of semantically flawed functions. In stark contrast, state-of-the-art learning-based baselines like DeGPT and SK2Decompile struggle severely with semantic hallucinations, achieving at most 48.29\% equivalence.
In summary, this paper makes the following contributions:

\begin{itemize}
  \item \textbf{Function-Isolated Recompilation and Validation under Global Consistency.}
  To address the fragility of whole-binary decompilation and the lack of function-level test oracles, we propose a function-isolated yet globally consistent recompilation paradigm. \sys restricts the LLM's reasoning space to single functions while strictly binding each to a unified compilation context to prevent definition conflicts. Furthermore, \sys hot-swaps the individually compiled function back into the unmodified binary via \emph{in-situ substitutable execution}, repurposing the program's native test suite as a robust and scalable behavioral validator.

\item \textbf{Fine-Grained Runtime Feedback via Breakpoint-Matched Differential Tracing.}
We introduce a two-stage dynamic feedback mechanism synergizing AddressSanitizer (ASan) with Breakpoint-Matched Differential Tracing (BP-Diff). While ASan swiftly isolates explicit memory violations, \sys relies on BP-Diff to expose deep, non-crashing semantic errors. By comparing control flow, registers, and memory states at matched breakpoints between original and substituted executions, \sys generates instruction-level evidence of behavioral divergences. This structured feedback directly guides the LLM to pinpoint root causes and surgically repair subtle logic defects.

\item \textbf{Effectiveness on real-world utility suites.}
We systematically evaluate \sys on complex, real-world binaries compiled from GNU \texttt{coreutils} and \texttt{binutils}. Our evaluation demonstrates that \sys achieves unprecedented function-level compilability and behavioral equivalence. By validating behavioral consistency against the official test suites, we show that \sys significantly outperforms both industry-standard rule-based decompilers and recent LLM-based refinement pipelines, reliably functioning even in challenging stripped binary scenarios.

\end{itemize}







\section{Background and Related Work}




Modern software assurance and modernization pipelines rely heavily on source-driven tooling, such as graph-centric mining (e.g., Joern~\cite{Joern2014}), pattern engines (e.g., Semgrep~\cite{Semgrep}), and automated source-level fuzzing. Furthermore, evolving legacy systems in critical domains (e.g., industrial control, finance) requires rigorous capabilities like cross-architecture porting and behavior-preserving hardening. 
Because these automated workflows strictly demand compilable and executable-equivalent code, relying purely on pseudocode with limited readability remains insufficient. Existing decompilation and refinement efforts generally approach this challenge from two distinct angles: optimizing for human readability, or attempting to repair code for usability.


\subsection{Decompilation for Readability and Local Semantics}
Traditional static decompilers such as Hex-Rays~\cite{idapro}, Ghidra~\cite{ghidra}, and RetDec~\cite{retdec} reconstruct syntax trees, control-flow graphs, and data-flow relationships to translate machine code into structured pseudocode. To further improve the structural fidelity of these tools, advanced compiler-aware structuring algorithms like SAILR~\cite{basque2024sailr} explicitly invert goto-inducing optimizations to recover control flows that closely mirror the original source code. 

Recent learning-based decompilation and refinement methods build on top of these tools by treating source recovery as a data-driven generation problem. One line of work focuses on whole-function lifting, using neural models or LLMs to produce decompiled code that is more readable, idiomatic, and closer to human-written sources. This includes early end-to-end neural decompilers such as Coda~\cite{fu2019coda} and Neural Decompilation~\cite{Omer2019NeuralDecompilation}, and later LLM-based systems (spanning methods based on prompt engineering, workflow design, or supervised fine-tuning) including LLM4Decompile~\cite{Tan2024llm4decompile}, ICL4Decomp~\cite{Wang2025icl4decomp}, CodeInverter~\cite{Liu2025codeinverter}, DeGPT~\cite{Hu2024DeGPT}, SALT4Decompile~\cite{Wang2025salt4decompile}, and SK2Decompile~\cite{Tan2025sk2decompile}.

Another line targets \emph{local semantic recovery}---types, symbols, and function signatures---to improve the quality of annotations for individual functions or regions, using analysis-driven type inference~\cite{Lee2011tie,Noonan2016retypd,Chen2020cati,Zhang2021osprey,Bosamiya2025trex,zhu2024tygr} and learning-based name and symbol prediction~\cite{He2018debin,Lacomis2019dire,Nitin2021direct,Chen2022dirty,Xu2023lmpa,Xie2024resym,Chen2025recopilot}. Together, these approaches significantly enhance the readability of decompiler output for human analysts.

However, these methods still fall short of our goal. They primarily optimize for readability and local semantic plausibility, and only rarely enforce that the resulting code can be compiled by standard toolchains and behaves equivalently to the original when executed inside the binary. Our work instead focuses on pseudocode translations that are explicitly required to compile and to perform as expected under real-world execution environments.

\subsection{Decompilation for Usability}
Beyond mere readability, a parallel research direction actively repairs decompiled pseudocode to guarantee both compilation success and strict behavioral equivalence. 

Several efforts attempt to improve decompilability without executing the code, relying instead on static pattern matching and retrieval. For instance, FidelityGPT~\cite{Zhou2026fidelitygpt} employs retrieval-augmented generation to detect and correct decompilation-induced artifacts. Yet, its reliance on limited predefined rules and retrieved historical patterns makes it brittle, restricting its ability to generalize to novel problems or highly obfuscated structures. Ultimately, without dynamic execution feedback, such purely static repair methods cannot validate their predictions against the actual native machine state, leaving deep semantic errors undetected.
Recognizing the limitations of purely static analysis, other methods incorporate execution or symbolic feedback into the repair loop. DeCLLM~\cite{Wong2025decllm} employs a systematic pipeline that uses test cases, AddressSanitizer, and fuzzing to detect semantic inconsistencies. However, because DeCLLM tests and repairs at the whole-file level, it suffers from severe \textit{error entanglement}. Long-propagated errors cascading across multiple functions create an intractable search space, drastically restricting its scalability on complex, real-world binaries. Alternatively, D-LIFT~\cite{Zou2025d-lift} incorporates compilation and symbolic execution into a backend feedback loop to guide RL fine-tuning. Unfortunately, its heavy reliance on symbolic execution scales poorly, often resulting in timeouts on large binaries and failing on complex memory models (e.g., double pointers). Furthermore, to make symbolic execution tractable, D-LIFT uses crude approximations for external function calls, preventing it from capturing deep runtime semantics and realistic control-flow dependencies. 

In summary, while recent methods strive to improve decompilation usability, they are individually bottlenecked by rigid static patterns, macro-level error entanglement, or unscalable symbolic constraints. In contrast, \sys integrates iterative, feedback-driven refinement within realistic native execution contexts. By explicitly isolating error scopes at the function level and leveraging fine-grained dynamic tracing, \sys systematically overcomes these boundaries, achieving both compilation success and strict behavioral equivalence on complex, real-world binaries.



\section{Problem Statement and Challenges}
\label{sec:problem}

\subsection{Problem Definition}

\textit{Setting.}
Let $B$ be a real-world binary comprising a set of functions $F$, and let $P = \{p_f\}$ denote the set of functions obtained by decompiling $B$. These pseudocode functions may fail to compile or, even if compilable, may diverge semantically from the behavior implemented in $B$.
Our goal is to recover a set of functions $T = \{t_f\}$ such that, for each function $f$:

\begin{itemize}
  \item \emph{Compilation Success:} $t_f$ compiles successfully with a standard compiler (e.g., \texttt{gcc}/\texttt{g++}) under an appropriate compilation context; and
  \item \emph{Behavioral equivalence:} when $t_f$ is substituted for the corresponding function in the original binary and invoked with the same inputs under the same program state, it exhibits the same observable behavior (outputs, side effects, and error conditions).
\end{itemize}

Although formulated at the function level, achieving verifiable equivalence for each individual function is an indispensable stepping stone toward whole-program recovery. Systematically accumulating these localized successes reconstructs the binary's execution logic, laying the groundwork for standalone file-level decompilation.

\subsection{Key Challenges}

\subsubsection{C1: Complexity of Globally Consistent Recompilation}
Raw decompiled pseudocode naturally lacks the definitions for custom types, global variables, and callee prototypes required for successful compilation. Naively prompting LLMs to hallucinate these missing dependencies on a per-function basis inevitably introduces fatal type and symbol inconsistencies across the binary. Therefore, the core complexity lies in achieving local compilation success for individual functions while strictly preserving global consistency to prevent conflicts during file-level composition.

\subsubsection{C2: Constructing Function-Level Equivalence Oracles}

Even after $t_f$ compiles successfully, verifying whether its runtime behavior matches the original $f$ remains challenging. Manual construction of test cases for individual functions is labor-intensive and often inadequate for complex real-world binaries. As discussed earlier, automated techniques relying on static analysis or symbolic execution encounter fundamental scalability and state-modeling bottlenecks. 
Consequently, the fundamental challenge is establishing a scalable verification mechanism for isolated functions that depend on global state, without resorting to expensive environment modeling or manual tests.

\subsubsection{C3: Localizing Behavioral Divergences for Repair}
Merely detecting that a translated function $t_f$ behaves differently from the original $f$ is insufficient for repair—existing methods yield signals that remain too coarse for error localization. Symbolic execution and SMT solvers may produce counterexamples or complex paths, but these outputs seldom indicate which code segments caused the divergence. Moreover, I/O operations at the binary level smear behavioral effects across stack frames, obscuring the root cause.
The core challenge, therefore, is to obtain feedback rich enough to pinpoint where and how $t_f$ deviates from $f$, thereby enabling an automated repair loop to operate effectively and at low cost.


\section{System Design: \sys}

\subsection{Architecture Overview}
\label{sec:overview}

To address the challenges, we present \textbf{\sys}, a feedback-driven framework that translates decompiler pseudocode into source code that can be compiled, executed in the native environment, and behaves equivalently to the original binary. The architecture comprises three modules.

\textbf{(1) Compile Functions with Consistent Minimal Context.}
Rather than allowing the LLM to hallucinate missing dependencies, \sys first establishes a minimal yet globally consistent compilation foundation. It extracts necessary type definitions, external function prototypes, and global variables directly from the binary using Hex-Rays' analysis. By anchoring the LLM's repair process to this extracted context ($\Gamma_f$), the system ensures that the locally repaired function can be successfully compiled into a standalone dynamic library without introducing cross-function symbol or type conflicts.

\textbf{(2) In-Situ Substitutable Execution as an Equivalence Oracle.}
To address \textbf{C2}, \sys introduces a practical, execution-based oracle for behavioral equivalence. This approach enables each individually compiled function to be hot-swapped back into the unmodified binary and executed in situ, thereby supporting local function replacement under globally consistent runtime conditions.
By reusing the original program’s test harness and execution environment, \sys effectively turns the binary’s own execution into a reliable equivalence validator. Any behavioral discrepancies observed during these runs directly indicate that the substituted function has not yet achieved full semantic alignment with the original.

\textbf{(3) Runtime Feedback–Guided Localization and Repair.}
When the substituted function fails the official test suite, \sys translates these coarse failures into fine-grained, actionable feedback. The diagnostic pipeline operates in two stages: it first intercepts explicit memory-safety violations using runtime sanitizers. If no crashes occur, it triggers Breakpoint-Matched Differential Tracing (BP-Diff) to align and compare execution states (i.e., control flow and variable values) against the original binary's ground truth. These structured diagnostics systematically guide the LLM in surgical, iterative repairs until the translated function precisely matches the original behavior.

\begin{figure*}[!t]
\centering
\includegraphics[width=\textwidth]{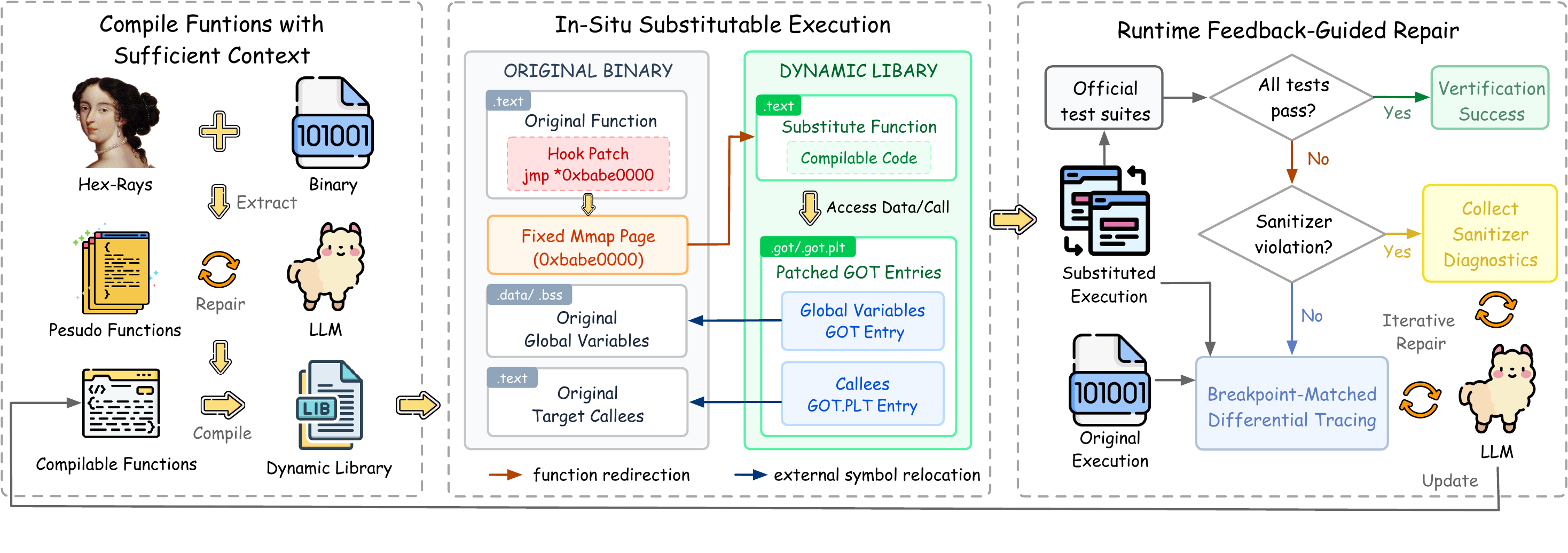}
\caption{Overview of the \sys workflow. 
(1) \textbf{Compile functions with sufficient context}: extract dependencies from the original binary and decompiled pseudocode, applying compiler-guided LLM repair to build compilable functions as dynamic libraries. 
(2) \textbf{In-situ substitutable execution}: seamlessly execute substitute functions inside the original binary using hook patches for function redirection and patched GOT entries for external symbol relocation. 
(3) \textbf{Runtime feedback-guided repair}: evaluate the substituted binary with official test suites. Failures trigger sanitizer diagnostics or \textbf{breakpoint-matched differential tracing}, providing fine-grained feedback for the LLM to iteratively repair the function.}
\label{fig:workflow}
\end{figure*}

Together, these modules instantiate a tight, iterative loop: compilation, in-situ substitution, and runtime feedback–guided repair, as summarized in Figure~\ref{fig:workflow}.

\subsection{Compile Functions with Sufficient Context}

To ensure successful compilation of functions, \sys reconstructs a minimal context for each target—including necessary type definitions, global variables, and callee prototypes. It then invokes a standard compiler on the unit. If compilation fails, the resulting diagnostics are fed back to the LLM, which repairs missing or inconsistent declarations and adjusts the translation accordingly. This process iterates until the function compiles successfully, as illustrated in the first step in Figure~\ref{fig:workflow}.

\subsubsection{Context Reconstruction}
\label{subsubsec:context}

To construct a minimal compilable context for each function, \sys integrates type dependencies, call relationships, and global variable references into a coherent whole. The reconstruction begins by collecting all decompiled functions from the binary and systematically analyzing their interdependencies through two complementary processes.

\textbf{Type Dependency Resolution.} To ensure the generated code is compilable, \sys constructs a comprehensive header file by consolidating decompiler-provided definitions (e.g., \texttt{defs.h}), platform-specific headers, and extracted type declarations. To minimize potential compilation errors, the framework systematically resolves inter-type dependencies to ensure all definitions are emitted in a valid order that satisfies the declaration-before-use requirement. By organizing these components into a unified context, \sys prevents issues related to incomplete types or missing declarations while maintaining global consistency across the binary. The concrete dependency resolution and type extraction mechanism is detailed in Section~\ref{sec:global_context_impl}.

\textbf{Global Context Integration.} To enable standalone compilation, \sys then reconstructs the execution context for each function by extracting function call prototypes to build a complete call graph and detecting global variable references to introduce necessary declarations. The resulting directed call graph is topologically ordered to guarantee dependency-safe code emission. This integrated approach produces self-contained functions that compile independently while maintaining linkability under global consistency constraints.

\subsubsection{Compiler-Guided Correction and Shared-Library Generation}
\label{subsubsec:llm_repair}
Despite accurate context reconstruction, a subset of decompiled functions still fail to compile. While standard headers resolve basic macros, they remain insufficient for more complex failures such as SIMD initialization mismatches or undeclared intrinsics. Residual artifacts frequently manifest as semantic type errors, including invalid pointer conversions, illegal pointer arithmetic, or references to undeclared internal primitives.

To resolve these compilation barriers, \sys employs a compiler guided correction loop. In this process, compilation diagnostics are structured as feedback to guide an LLM in proposing targeted fixes, which may include idiom rewriting, type reconciliation, and conflict resolution. This iterative process continues until each function reaches a stable, compilable state within its context $\Gamma_f$.
Each successfully corrected function is then compiled into a dynamic library (e.g., \texttt{.so} or \texttt{.dll}). This packaging enables dynamic linking of individual functions back into the original binary, establishing the foundation for subsequent in-situ substitution and behavioral equivalence verification.

\subsection{In-Situ Substitutable Execution}
\label{subsec:substitutable_exec}

To validate behavioral equivalence at function granularity, each function must be executed in situ, that is, within a running instance of the original binary. This ensures that any observed behavioral discrepancies can be attributed solely to the substituted function.
A central challenge is to execute the substituted function without altering other modules, while maintaining correct interaction with global state and callee functions. To this end, \sys establishes precise address and symbol mappings and performs bidirectional runtime relocation, integrating the substitute seamlessly into the unmodified binary.

\subsubsection{Address Mapping and Symbol Correspondence Construction}
\label{subsubsec:addr_map_design}

\begin{figure}[!t]
  \centering
  \includegraphics[width=0.99\linewidth]{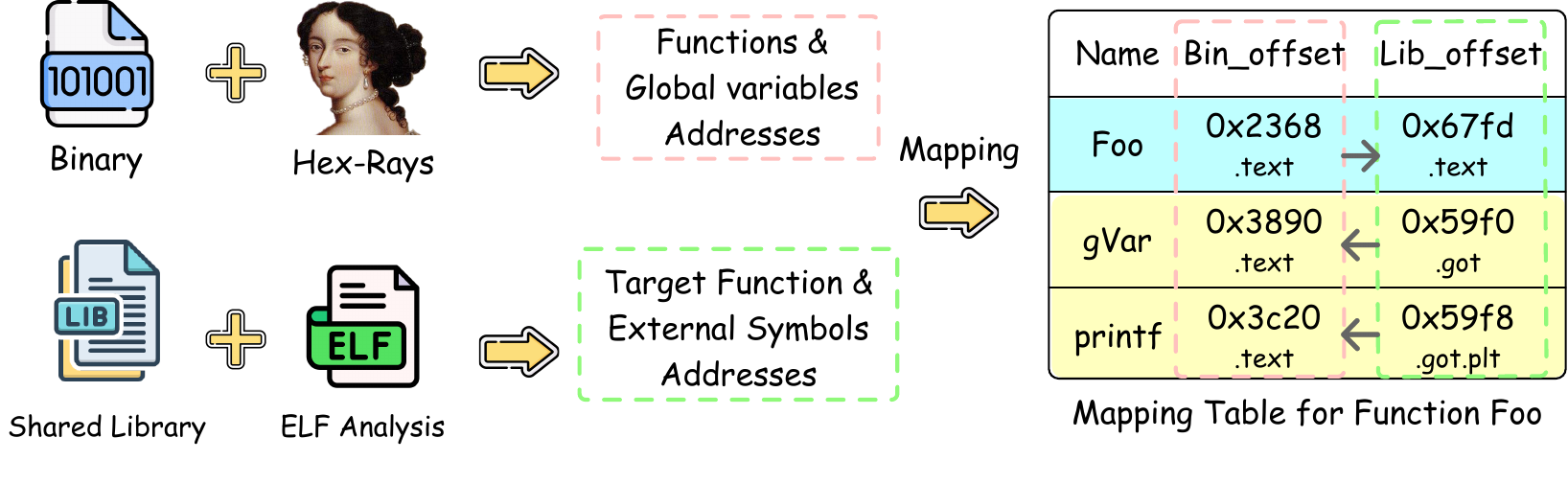}
  \caption{Construction of the address mapping table. \sys aligns symbols from the original binary with those in the compiled dynamic module to enable bidirectional runtime relocation.}
  \label{fig:mapping_table}
\end{figure}

During the static analysis phase, \sys constructs a unified mapping table to align the substituted function with its external dependencies. As illustrated in Figure~\ref{fig:mapping_table}, this table securely connects symbols in the compiled dynamic module to their exact memory counterparts in the original binary.

To build this table, \sys extracts the offsets of the original functions and global variables from the target binary using Hex-Rays analysis. Concurrently, it performs standard ELF analysis on the compiled dynamic module to pinpoint the offsets of the substitute function and its external symbol references.
By matching these two analysis streams, \sys constructs a comprehensive mapping table consisting of \( \langle \texttt{Name}, \texttt{Bin\_offset}, \texttt{Lib\_offset} \rangle \) tuples. For the target function itself (e.g., \texttt{Foo} in Figure~\ref{fig:mapping_table}), the tuple links its original entry address in the binary's \texttt{.text} segment (\texttt{Bin\_offset}) to the new implementation's entry address in the dynamic module (\texttt{Lib\_offset}). 
Crucially, to enable bidirectional interaction, \sys creates analogous tuples for all external references accessed by the substitute function. As shown in the figure's mapping table, global variables (e.g., \texttt{gVar}) accessed via the dynamic module's \texttt{.got} are mapped back to their original absolute offsets in the binary. Similarly, external callee functions (e.g., \texttt{printf}) routed through the module's \texttt{.got.plt} are mapped directly to their corresponding target addresses in the original binary's \texttt{.text} segment. 

This comprehensive mapping table serves as a deterministic binding contract for the runtime engine, enabling complex dynamic substitution without requiring on-disk rewriting or re-linking of the original binary.

\subsubsection{Bidirectional Runtime Relocation}
\label{subsubsec:runtime_bind_design}

When the original binary starts, a lightweight relocation engine is automatically loaded by the dynamic linker. To account for Address Space Layout Randomization (ASLR), the engine first dynamically resolves the runtime load base addresses of both the original binary and the compiled dynamic module. Using these base addresses combined with the pre-computed mapping table, it enables substitutable execution through a two-way relocation process, as illustrated in the \textit{In-Situ Substitutable Execution} phase of Figure~\ref{fig:workflow}.

First, the engine performs \emph{function redirection} to route execution into the substitute. Using the original offset recorded in the mapping table, it locates the function's entry point within the binary and dynamically overwrites it with a hook patch. As shown in the figure, this patch utilizes a fixed \texttt{mmap} page as a trampoline to redirect control flow directly to the substitute function's corresponding offset in the dynamic module. This post-load patching modifies only the targeted function, leaving the surrounding executable environment intact.
Second, the engine conducts \emph{external symbol relocation} to grant the substitute access to the original execution context. When the substitute function accesses data or calls external dependencies, it natively relies on its own \texttt{.got} and \texttt{.got.plt} tables. To synchronize this state, the engine queries the previously constructed mapping table to retrieve the precise offsets of these external symbols. Combining these with the resolved base addresses, it calculates absolute runtime addresses and directly patches the substitute's GOT entries. As depicted in the figure, this accurately routes data accesses and subroutine calls back to the original global variables and target callees.

Beyond direct dependencies, this shared execution context naturally accommodates indirect calls, such as virtual functions and callbacks. Since decompiled pseudocode represents these invocations as raw memory dereferences derived from passed arguments or global structures rather than introducing new external symbols, they require no explicit patching. By operating directly within the original address space, the substitute natively utilizes unmodified pointer arguments and accesses correctly relocated dispatch tables (e.g., vtables). Thus, runtime dereferences naturally fetch valid targets. Applied in-memory post-load (§~\ref{sec:relocation_engine}), these relocations ensure transparent in-situ execution.

\subsubsection{Advantages over Static Binary Rewriting}
\label{subsubsec:advantages_static_rewriting}

While prior works like PRD~\cite{Reiter2022PRD} target vulnerability patching by replacing recompensable functions, they rely on static on-disk rewriting designed for permanent fixes. To link external symbols, PRD intrusively alters the substituted function's prototype to pass memory addresses as arguments and injects fragile, architecture-specific inline assembly. This distortion of standard C semantics creates a barrier for LLMs that rely on clean code contexts. In contrast, by operating entirely in memory via explicit GOT hijacking, \sys allows the LLM to generate and interact exclusively with standard, unmodified C code. 
Furthermore, our approach exhibits superior cross-platform scalability. Static rewriters are typically tightly coupled to specific binary formats through complex ELF section manipulation. By abstracting the integration into runtime pointer fixups, \sys's strategy seamlessly generalizes to other operating systems. Adapting to Windows (PE) is conceptually identical: the substitute compiles as a DLL, while dynamic hooking (e.g., Microsoft Detours) and RVA resolution against the host's load base handle redirection. This provides a versatile foundation for verifiable decompilation across diverse ecosystems.

\subsection{Runtime Feedback-Guided Repair}
\label{subsec:runtime_feedback_design}

Once in-situ substitution is enabled, \sys evaluates the modified binary using official test suites. If all tests pass, the function is successfully verified. Otherwise, it initiates a targeted diagnostic pipeline to pinpoint divergence root causes. As depicted in Figure~\ref{fig:workflow}, while basic memory violations are quickly intercepted via sanitizer diagnostics, the core repair engine relies on breakpoint-matched differential tracing. This technique systematically exposes deep semantic defects by comparing original and substituted executions. Together, these fine-grained runtime signals provide the LLM with structured feedback for iterative repair until verification succeeds.

\subsubsection{Sanitizer Diagnostics and Repair}
\label{subsubsec:asan_design}

Sanitizers offer direct detection of memory errors that may cause execution crashes or deep semantic divergence. To isolate errors specifically introduced by the decompilation and LLM-rewrite processes, \sys employs an asymmetric partial sanitization strategy. 
During the recompilation of the substituted function, we explicitly enable AddressSanitizer (ASan)~\cite{Konstantin2012AddressSanitizer} instrumentation (\texttt{-fsanitize=address}). At runtime, the corresponding ASan runtime library is globally injected into the host process (as detailed in §\ref{sec:relocation_engine}). This asymmetric setup ensures that the ASan runtime intercepts the host program's global memory allocators (e.g., \texttt{malloc} and \texttt{free}) to place protective redzones around all native heap objects, while memory access checks are strictly enforced only within the LLM-generated substitute. 

When ASan detects a violation, a targeted patching phase guides the LLM to resolve the issue using structured diagnostics. The primary objective of this phase is to eliminate decompilation-induced artifacts that crash the dynamic oracle and halt verification, such as uninitialized variables leading to miscalculated loop bounds. 

\subsubsection{Breakpoint-Matched Differential Tracing and Repair}
\label{subsubsec:bp_design}\label{subsubsec:loop_design}

When a substituted function fails verification without triggering sanitizer violations, \sys initiates \emph{breakpoint-matched differential tracing} to expose subtle semantic divergences. For each failing input, \sys executes the binary in both its original and substituted states under identical conditions, leveraging the GNU Debugger (GDB)~\cite{GDB} to trace and compare internal execution states.

\begin{figure}[!t]
  \centering
  \includegraphics[width=1\linewidth]{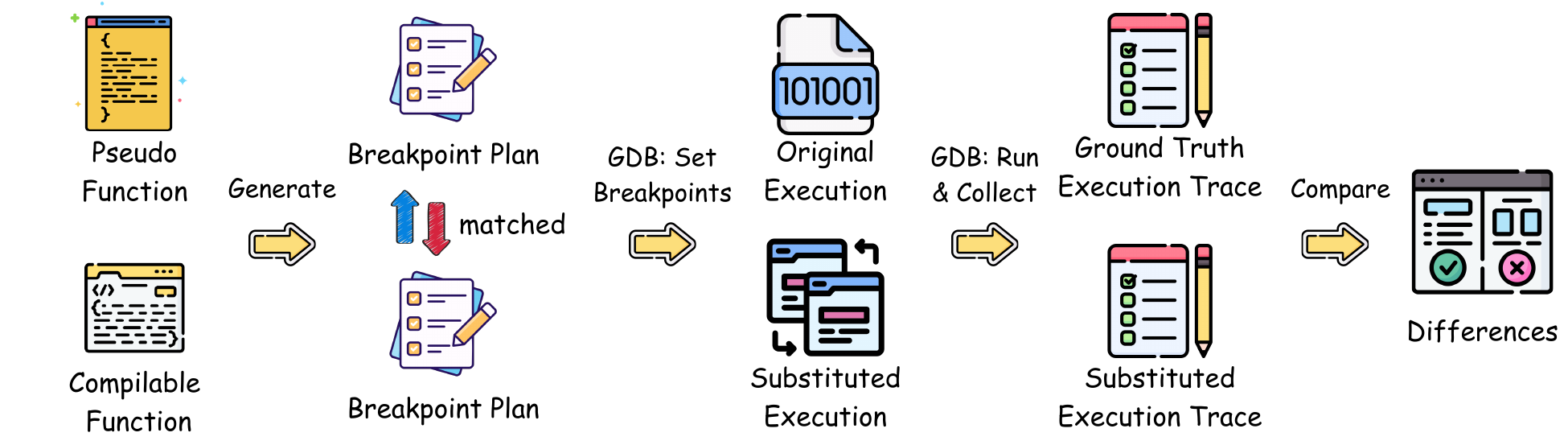}
  \caption{Workflow of breakpoint-matched differential tracing.}
  \label{fig:diff_tracing}
\end{figure}

\textbf{Generating matched breakpoints.} 
As illustrated in Figure~\ref{fig:diff_tracing}, \sys first determines where to monitor execution by jointly analyzing the compilable function and its raw pseudocode counterpart. Because the compilable function generally maintains statement-level isomorphism with the pseudocode, a structural diff seamlessly aligns corresponding statements. Based on this alignment, \sys generates a \emph{breakpoint plan} specifying positionally matched source locations across both representations, alongside the specific variables to track at each site. For example, breakpoints at assignments focus on tracking their left-hand values, while locations without specific variables serve as control-flow markers.

\textbf{Trace collection.} 
Guided by the breakpoint plan, \sys configures GDB to attach to the target process, compute the appropriate runtime code addresses from the module base and install the matched breakpoints. During execution, each breakpoint hit records a control-flow marker and captures the normalized values of designated variables. Executing the native binary yields a \emph{ground truth execution trace}, while running the binary with the dynamic module yields a \emph{substituted execution trace} collected under the identical plan.

\textbf{Trace comparison and iterative repair.} 
To identify discrepancies, trace snapshots are first aligned at function-call granularity using invocation indices and call depths (to disambiguate recursive or multiple invocations). Within each matched call, a comparator performs breakpoint-level alignment between the substituted trace and the ground truth to flag behavioral differences, such as missing/extra breakpoint hits or variable value mismatches. This analysis produces a structured diff recording the breakpoint location, source context from the substituted function, divergent execution paths or variable values, and the last successfully matched site to bound the suspected error region. 
These normalized differences are then provided to the LLM to guide targeted semantic edits. 
Following each modification, the updated function is rebuilt into a dynamic module and subjected to further paired executions until no failed tests remain or a predefined repair budget is exhausted.

\section{Implementation}




\subsection{Function-Level Compilation Pipeline}
\label{sec:global_context_impl}
Beyond the conceptual design, we implement a concrete pipeline for compiling pseudocode functions in practice. Our implementation targets Linux (x86\_64) with Hex-Rays, chosen because prior work~\cite{Gao2025Decompilebench} indicates that its decompiler output is closest to compilable and practically usable, and because it can be driven in batch via noninteractive scripts.



\textbf{Type Extraction and Dependency Resolution.} The pipeline begins by extracting local type declarations, global variables, and external function prototypes for each function. For unstripped binaries, we directly retrieve precise type information through Hex-Rays' Python APIs. For stripped binaries, where symbolic metadata is absent, we integrate a structure recovery algorithm ~\cite{anonymous_circle} to reconstruct struct and union layouts from raw binary artifacts. 
We leverage IDA's provided \texttt{defs.h} header to map standard Hex-Rays primitives (e.g., \texttt{\_DWORD}). For remaining non-standard idioms that are unresolved by the standard header, such as opaque \texttt{JUMPOUT(0x...)} control-flow artifacts, we employ a symbol-driven heuristic pass. This pass queries the decompiler's symbol table to map the raw hex addresses back to valid function call signatures (e.g., transforming \texttt{JUMPOUT(0x2885);} into \texttt{quotearg\_buffer\_restyled\_cold();}).
All extracted and recovered types are then organized into a Type Dependency Graph (TDG). We classify dependencies into three categories: (1) typedef dependencies, (2) function and pointer dependencies, and (3) user-defined type (UDT) member dependencies. 
A topological sort of the TDG enforces a dependency-safe emission order, where forward declarations, enums, and typedefs strictly precede complete UDT definitions, effectively preventing compilation errors from circular dependencies or incomplete types. The resulting normalized types are then consolidated into a unified header to maintain global consistency across functions.

\textbf{Compilation-guided repair.}
Each function is then compiled as a dynamic library. Compiler diagnostics are captured as structured records that include the error kind, source location, code span, and a context snippet. These records are fed into a bounded repair loop that invokes the LLM to refine the function body or adjust surrounding declarations, after which the function is rebuilt. The process repeats until compilation succeeds or the budget is exhausted.



\subsection{Relocation Engine Implementation}
\label{sec:relocation_engine}

This section details the system-level implementation of the bidirectional runtime relocation mechanism introduced in §\ref{subsec:substitutable_exec}. The relocation engine is compiled as a standalone helper library and injected into the target process alongside the LLM-repaired dynamic library.

\textbf{Injection and Base Resolution.} 
To ensure all redirections are established before the original binary executes, the engine is injected via \texttt{LD\_PRELOAD}. During the automated testing phase that requires memory safety diagnostics (§\ref{subsubsec:asan_design}), the ASan runtime library (e.g., \texttt{/usr/lib/.../libasan.so}) is co-injected alongside the relocation engine.
Its constructor hooks execute strictly before the host program's \texttt{main()} function. Upon initialization, the engine parses \texttt{/proc/self/maps} to dynamically retrieve the memory load bases of both the host binary (\texttt{bin\_base}) and the injected dynamic library (\texttt{lib\_base}). For debugger-assisted workflows, the engine can optionally suspend the process at this stage, exposing a resume mechanism that allows GDB to attach and set breakpoints deterministically.

\begin{figure}[!t]
  \centering
  \includegraphics[width=0.95\linewidth]{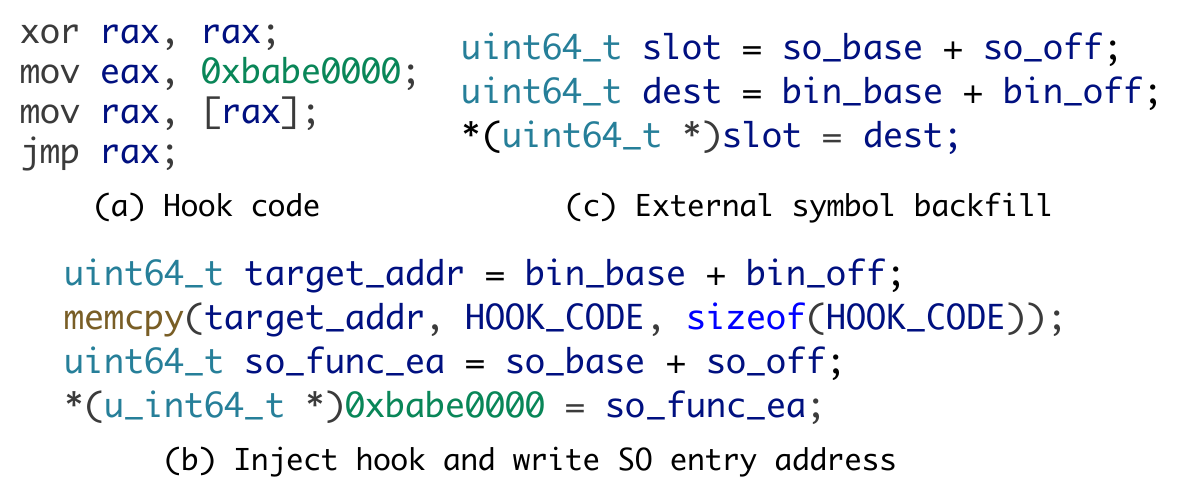}
  \caption{Runtime bidirectional relocation by the relocation engine.}
  \label{fig:hook}
\end{figure}

\textbf{Control-Flow Hijacking Mechanics.} 
To redirect execution to the substituted function, the engine temporarily elevates the memory permissions of the host binary's \texttt{.text} page to read/write/execute. It then overwrites the original function's entry point with a compact control-transfer stub (trampoline), as shown in Figure~\ref{fig:hook}(b). Specifically, the injected hook consists of a four-instruction sequence that loads a 64-bit callee pointer from a dedicated memory cell (e.g., mapped at \texttt{0xbabe0000}) and performs a tail jump to the dynamic library (Figure~\ref{fig:hook}(a)). The page is subsequently restored to read/execute permissions to maintain stability. 


\subsection{Activating Breakpoints at Runtime}

For the Set Breakpoints stage in step (3) of Figure~\ref{fig:workflow}, a key challenge lies in bridging the gap between the source-level breakpoint plan, which refers to locations in the compilable or pseudocode, and the binary-level breakpoints that the debugger must install at concrete machine instruction addresses. To apply the same breakpoint plan consistently across both the original and substituted executions, we resolve each planned site to concrete binary addresses at load time and install instruction-level traps.

Concretely, \sys relies on two minimal and stable mapping anchors:
(i) a binary–source mapping for the compiled function, and
(ii) a binary–pseudo mapping for the decompiled pseudocode function.
Hex-Rays provides a bidirectional correspondence between decompiled pseudocode and the underlying assembly instructions, including the concrete storage locations (registers or stack slots) of pseudocode variables. We leverage this mapping to translate pseudocode-level breakpoint sites into instruction addresses in the original binary. Similarly, when recompiling the translated function into a dynamic library, we enable debugging information (e.g., using \texttt{-g}) to establish line-to-instruction mappings, while also leveraging Hex-Rays' existing source-to-assembly correspondences to maintain consistent breakpoint alignment across both execution environments.

Using these two mappings, the debugger attaches to both the original and substituted processes and translates each logical breakpoint identifier from the plan into a process-specific absolute address. As a result, the same logical site, defined at the source or pseudocode level, is consistently instrumented in both executions. This allows us to collect directly comparable runtime state snapshots from the original and the translated function.

\section{Evaluation}

\subsection{Experimental Setup}
\label{sec:exp-setup}

\noindent\textbf{Environment.}
All experiments were conducted on Ubuntu~22.04. We used Hex-Rays~9.1, Python~3.12, and GDB~12.1. All translated functions were compiled with \texttt{g++~11.4.0}, with symbol stripping disabled so that debugging information remained available. The evaluations ran on a server equipped with a 96-core \texttt{Intel(R) CPU @ 3.00GHz} and an \texttt{NVIDIA Tesla V100-PCIE-32GB} GPU.

\noindent\textbf{Models used.}
The repair and refinement components were evaluated with three LLMs accessed via API:
\textbf{GPT-5.1}, \textbf{DeepSeek-V3}, and \textbf{Qwen3-Coder-480B-A35B-Instruct}.
For DeepSeek-V3 and Qwen3-Coder-480B-A35B-Instruct, we set \texttt{temperature = 0.2}, \texttt{top\_p = 0.95}, and \texttt{max\_tokens = 8192}. For GPT-5.1, which does not expose these controls, we use the default configuration. To balance repair effectiveness with computational cost, we cap the number of LLM repair iterations to at most \texttt{5} for compilation errors and \texttt{3} for runtime errors per function. 
All prompts share a fixed scaffold (see Appendix~\ref{subsec:repair_executor}).

\noindent\textbf{Runtime settings.}
We enable AddressSanitizer (ASan) in the test harness to surface memory errors. Execution runs in a restricted runner with a per-test-case timeout of \texttt{120\,s}. Function replacement is performed at \emph{function} granularity via our loader. Crashes, timeouts, or sanitizer violations are all treated as failures.

\subsection{Datasets and Metrics}
\label{sec:metrics-benchmarks}

\textbf{Datasets.}
We construct our datasets from two established Linux C suites, \textit{coreutils-9.5} and \textit{binutils-2.45}. 
For each suite, we build both \emph{unstripped} and \emph{stripped} reference binaries using the standard build systems with GCC~\cite{FSFGCC} at the \texttt{-O2} optimization level. We then decompile these binaries to obtain the inputs to our pipeline—per-function pseudocode together with the \emph{compilation context} required for standalone compilation. 
The resulting datasets comprise \textbf{108} Coreutils binaries and \textbf{12} Binutils binaries. For Coreutils, the unstripped versions average \textbf{98} functions per binary (\textbf{47} lines/function), while the stripped versions average \textbf{125} functions per binary (\textbf{59} lines/function). For Binutils, the unstripped versions average \textbf{1{,}587} functions per binary (\textbf{86} lines/function), while the stripped versions average \textbf{1{,}813} functions per binary (\textbf{94} lines/function). These aligned artifacts form the basis of our evaluation.

To assess correctness, we rely on each suite's official test suites. For \emph{GNU Coreutils}, we run the provided regression tests (shell/Perl harnesses), capture \texttt{stdout}/\texttt{stderr} and exit codes, and deem a run correct when the observed outputs match the oracle. For \emph{GNU Binutils}, we use the DejaGnu \texttt{runtest} harness and derive success from its summaries by aggregating ``expected passes'' and ``unexpected failures.''

\textbf{Metrics.}
We report two metrics---\emph{Compilation Success} and \emph{Behavioral Equivalence}---each evaluated at two granularities: function-level and file-level. 
At the \emph{function level}, \emph{Compilation Success} is the fraction of translated functions that compile, and \emph{Behavioral Equivalence} is the fraction that both compile and pass all associated tests. Note that \emph{Behavioral Equivalence} is strictly bounded by the coverage of the official test suites; functions not exercised by the test harness evade runtime validation. Nevertheless, our framework is inherently scalable to large, automated test suites, though the orthogonal challenge of test case generation remains beyond the scope of this paper.
At the \emph{file level}, success serves as an aggregated metric. \emph{Compilation Success} considers a binary successful only if all of its translated functions compile correctly. 
The \emph{Behavioral Equivalence} criterion requires every individual constituent function within the binary to pass the full test suite during its isolated in situ execution. This bottom-up design ensures that localized improvements cumulatively guarantee whole-binary correctness, paving the way for accurate binary-source recovery.

\begin{table*}[t]
\renewcommand{\arraystretch}{1.15}
\setlength{\tabcolsep}{4pt}
\caption{Compilation success (CS) and behavioral equivalence (BE) on Coreutils and Binutils (function/file levels).}
\label{tab:overall-results}
\centering
\small
\resizebox{\textwidth}{!}{%
\begin{tabular}{l l c c c c c c c c c c c c c c}
\toprule
\multirow{2}{*}{\textbf{Benchmark}} & \multirow{2}{*}{\textbf{Granularity}} 
& \multicolumn{2}{c}{\textbf{HexRays}} 
& \multicolumn{2}{c}{\textbf{Ghidra}} 
& \multicolumn{2}{c}{\textbf{DeGPT}} 
& \multicolumn{2}{c}{\textbf{SK2Decompile}} 
& \multicolumn{2}{c}{\textbf{DeCLLM}} 
& \multicolumn{2}{c}{\textbf{FidelityGPT}} 
& \multicolumn{2}{c}{\textbf{pcodetrans}} \\
\cmidrule(lr){3-4}\cmidrule(lr){5-6}\cmidrule(lr){7-8}\cmidrule(lr){9-10}\cmidrule(lr){11-12}\cmidrule(lr){13-14}\cmidrule(lr){15-16}
& &
\textbf{CS} & \textbf{BE} &
\textbf{CS} & \textbf{BE} &
\textbf{CS} & \textbf{BE} &
\textbf{CS} & \textbf{BE} &
\textbf{CS} & \textbf{BE} &
\textbf{CS} & \textbf{BE} &
\textbf{CS} & \textbf{BE} \\
\midrule
\multirow{2}{*}{Coreutils} 
& Function-level & 69.03\% & 65.98\% & 55.41\% & 52.00\% & 53.30\% & 48.29\% & 31.22\% & 24.61\% & -- & -- & 49.44\% & 47.04\% & \textbf{100\%} & \textbf{99.55\%} \\
& File-level     & 0\% & 0\%  & 0\%  & 0\%  & 0\% & 0\% & 0\% & 0\% & 70.37\% & 55.56\% & 0\%  & 0\% & \textbf{100\%} & \textbf{81.48\%} \\
\midrule
\multirow{2}{*}{\makecell[l]{Coreutils-\\stripped}} 
& Function-level & 61.68\% & 55.34\% & 49.40\% & 45.67\% & 48.72\% & 43.11\% & 15.78\% & 13.94\% & -- & -- & 34.29\% & 29.81\% & \textbf{99.12\%} & \textbf{96.84\%} \\
& File-level     & 0\% & 0\% & 0\% & 0\% & 0\% & 0\% & 0\% & 0\% & -- & -- &  0\% & 0\% & \textbf{44.44\%} & \textbf{25.92\%} \\
\midrule
\multirow{2}{*}{Binutils} 
& Function-level & 80.12\% & 78.61\% & 43.14\% & 42.34\% & 39.31\% & 38.52\% & 21.30\% & 18.95\% & -- & -- & 51.83\% & 50.57\% & \textbf{100\%} & \textbf{99.89\%} \\
& File-level     & 0\% & 0\% & 0\% & 0\% & 0\% & 0\% & 0\% & 0\% & -- & -- & 0\% & 0\% & \textbf{100\%} & \textbf{75.00\%} \\
\midrule
\multirow{2}{*}{\makecell[l]{Binutils-\\stripped}} 
& Function-level & 74.65\% & 73.03\% & 40.39\% & 39.64\% & 40.05\% & 39.28\% & 8.69\% & 7.88\% & -- & -- & 36.81\% & 36.08\% & \textbf{99.15\%} & \textbf{98.38\%} \\
& File-level     & 0\% & 0\% & 0\% & 0\% & 0\% & 0\% & 0\% & 0\% & -- & -- & 0\% & 0\% & \textbf{16.67\%} & \textbf{25.00\%} \\
\bottomrule
\end{tabular}
}
\end{table*}

\subsection{Results and Ablation Studies}
\label{sec:results-ablation}

\subsubsection{Effectiveness of \sys}

To assess the end-to-end effectiveness of \sys, we evaluate its ability to produce compilable and behaviorally equivalent C source code from raw binary artifacts. Our evaluation compares \sys against both industry-standard decompilers and state-of-the-art LLM pipelines across unstripped and stripped versions of \textit{Coreutils} and \textit{Binutils}. By analyzing function- and file-level performance, we demonstrate that \sys significantly outperforms existing methods, with overall quantitative results summarized in Table~\ref{tab:overall-results}.

\textbf{Baselines Setup.}
We evaluate \sys against both industry-standard decompilers and recent LLM-based repair pipelines. For raw decompilers, we use Hex-Rays and Ghidra. 
For open-source LLM pipelines, we evaluate SK2Decompile, DeGPT, and FidelityGPT. 
Additionally, since DeCLLM~\cite{Wong2025decllm} lacks a public release, we include its officially reported results on the unstripped \textit{Coreutils} dataset for direct comparison. Likewise, we omit empirical comparisons with D-LIFT~\cite{Zou2025d-lift}, as its training pipeline and fine-tuned models currently remain completely closed-source, precluding any local replication.
To ensure a fair comparison, we provide Ghidra with a comparable minimal context by extracting callee prototypes and global variable declarations via cross-reference analysis, and supply a header to mitigate common Ghidra-specific artifacts.


Since LLM baselines typically output only isolated function bodies that rarely compile standalone, evaluating their raw outputs would yield trivially low success rates. To establish a rigorous upper bound, we deliberately equip these baselines with our systematically extracted compilation contexts. Specifically, both SK2Decompile and FidelityGPT operate on the enhanced Hex-Rays pseudocode and are paired with the Hex-Rays-derived context. For FidelityGPT, we faithfully reproduce its static two-stage retrieval-augmented generation (RAG) workflow (i.e., distortion annotation and subsequent correction), utilizing \texttt{deepseek-chat} as the backbone LLM and \texttt{text-embedding-ada-002} for retrieval. Meanwhile, for DeGPT, we apply its official \texttt{DeepSeek-V3} workflow to the aligned Ghidra pseudocode, wrapping its outputs in the corresponding Ghidra-derived context. Ultimately, all evaluated baselines rely exclusively on their respective static decompiler outputs and contexts, and are compiled directly without any iterative repair loops.

\textbf{Quantitative results and failure analysis.} 
\sys markedly outperforms all baselines across the four evaluated datasets. At the function level, it consistently achieves near-perfect behavioral equivalence, reaching 99.55\% on \textit{Coreutils} and 99.89\% on \textit{Binutils}. Scaling to the file level, where the compounding effect of minor semantic artifacts causes most baseline methods to completely fail (0\% success), \sys successfully recovers 81.48\% of \textit{Coreutils} binaries and 75.00\% of \textit{Binutils} binaries as fully executable equivalents. 

Stripping symbolic metadata expectedly degrades performance across all tools. However, \sys exhibits strong resilience, maintaining over 96\% function-level equivalence on both stripped datasets and remaining the sole pipeline capable of yielding functional whole-program binaries. Notably, the steep drop from function-level to file-level success on stripped datasets (e.g., 25.92\% on \textit{Coreutils-stripped}) underscores the compounding difficulty of whole-program recovery. Because file-level evaluation strictly demands that every constituent function be perfectly repaired, a single flawed function immediately cascades into a file-level failure. Furthermore, we evaluate \sys against DeCLLM, the only baseline specifically designed for whole-binary repair, using its reported unstripped \textit{Coreutils} dataset. \sys achieves a decisive advantage in both compilation (100\% vs. 70.37\%) and file-level equivalence (81.48\% vs. 55.56\%). This contrast demonstrates that our bottom-up strategy of enforcing rigorous dynamic verification at the function granularity before file-level integration is superior to directly operating at the macroscopic level.

Among traditional decompilers, Hex-Rays consistently outperforms Ghidra, yet their performance divergence is highly dataset-dependent. On \textit{Binutils}, Hex-Rays dramatically leads in behavioral equivalence (78.61\% vs. 42.34\%) because Ghidra's heuristics struggle with complex pointer topologies like BFD trees, causing severe type collapse. Conversely, this gap narrows significantly on \textit{Coreutils} (65.98\% vs. 52.00\%). In this environment-heavy project, Hex-Rays' tendency to over-restore system-level symbols (e.g., \texttt{struct passwd}) backfires, frequently triggering redefinition conflicts that suppress overall compilability.


Learning-based baselines introduce distinct error modalities. Optimized primarily for code readability, they inadvertently sacrifice compilability and semantic correctness, often degrading the raw outputs of their underlying decompilers. SK2Decompile plummets relative to Hex-Rays, falling to 24.61\% equivalence on \textit{Coreutils} and 18.95\% on \textit{Binutils}. This severe degradation stems from hallucinated data structures and arbitrarily altered symbol names, which break external linkages and render meticulously provided contextual information ineffective. 
While FidelityGPT exhibits fewer symbol hallucinations, its chunk-based correction introduces severe structural flaws when processing long functions. Uncoordinated local edits cause premature scope closures (e.g., extra \texttt{\}}) and mismatched cross-chunk variables, producing fundamentally uncompilable code plagued by unbalanced brackets. This fragility is particularly exacerbated on stripped binaries, yielding only 34.29\% compilation success on \textit{Coreutils-stripped}. 
Meanwhile, DeGPT exhibits a milder penalty, dropping to 48.29\% equivalence on \textit{Coreutils}. Although structurally cleaner, it consistently suffers from context-unaware typing and undeclared variables. Ultimately, these diverging failure modes underscore the absolute necessity of a context-aware, compiler-driven dynamic repair loop to achieve strict semantic equivalence.

\begin{table}[t]
\renewcommand{\arraystretch}{1.2}
\setlength{\tabcolsep}{3pt}
\caption{
Sequential debug-repair effectiveness on \textit{Coreutils} and \textit{Binutils}. 
``Comp.\ Succ.'' denotes the compilation success rate. 
``Runtime Failed'' indicates functions that compiled but failed official tests. 
The repair columns report the number (and percentage) of these failures successfully fixed.
}
\label{tab:debugrepair-coreutils-binutils}
\centering
\small
\resizebox{\columnwidth}{!}{
\begin{tabular}{l l c c c c}
\toprule
\textbf{Dataset} & \textbf{Model}
  & \makecell{\textbf{Comp.}\\\textbf{Succ.\%}}
  & \makecell{\textbf{Runtime Failed}\\\scriptsize(Post-Comp.)} 
  & \makecell{\textbf{+ ASan Fix}\\\scriptsize(Count \& \%)} 
  & \makecell{\textbf{+ BP-Diff Fix}\\\scriptsize(Count \& \%)} \\
\midrule
\multirow{3}{*}{Coreutils}
& DeepSeek\texttt{-}V3   & 100\% & 320 & 17 (5.31\%)  & 141 (44.06\%) \\
& Qwen3\texttt{-}Coder   & 100\% & 320 & 20 (6.25\%)  & 157 (49.06\%) \\
& GPT\texttt{-}5.1        & 100\% & 320 & 20 (6.25\%)  & 245 (76.56\%) \\
\midrule
\multirow{3}{*}{\makecell[l]{Coreutils-\\stripped}}
& DeepSeek\texttt{-}V3   & 90.72\% & 364 & 15 (4.12\%) & 90 (24.73\%) \\
& Qwen3\texttt{-}Coder   & 94.02\% & 397 & 41 (10.32\%) & 185 (46.60\%) \\
& GPT\texttt{-}5.1        & 99.12\% & 451 & 50 (11.09\%) & 239 (53.00\%) \\
\midrule
\multirow{3}{*}{Binutils}
& DeepSeek\texttt{-}V3   & 100\% & 153 & 8 (5.23\%)   & 78 (50.98\%) \\
& Qwen3\texttt{-}Coder   & 100\% & 153 & 8 (5.23\%)   & 84 (54.90\%) \\
& GPT\texttt{-}5.1        & 100\% & 153 & 9 (5.88\%)   & 122 (79.74\%) \\
\midrule
\multirow{3}{*}{\makecell[l]{Binutils-\\stripped}}
& DeepSeek\texttt{-}V3   & 90.68\% & 228 & 15 (6.60\%) & 72 (31.58\%) \\
& Qwen3\texttt{-}Coder   & 95.38\% & 326 & 47 (14.41\%) & 183 (56.13\%) \\
& GPT\texttt{-}5.1        & 99.15\% & 353 & 49 (13.88\%) & 208 (58.92\%) \\
\bottomrule
\end{tabular}
}
\end{table}

\subsubsection{Effect of Context on Compilation Success}
\label{subsubsec:ablation-context}
The ablation shows a clear monotonic relationship between context completeness and compilability across both unstripped and stripped binaries. As shown in Table~\ref{tab:ablation-context}, starting from \emph{pseudocode only}, compilation rates are extremely low (e.g., 0.51\% for unstripped \textit{Coreutils} and 0.19\% for its stripped counterpart). Introducing local type definitions removes many declaration-before-use and signature errors, raising success noticeably (e.g., to 9.58\% and 12.64\% on \textit{Coreutils}). The largest single gain consistently comes from adding callee declarations, which lifts rates substantially (up to 53.44\% and 67.80\% for unstripped binaries) and highlights the critical importance of cross-call type constraints. 
Combining all context sources (\emph{local types + callee + global}) achieves 69.03\% and 80.12\% on unstripped \textit{Coreutils} and \textit{Binutils}. Expectedly, on stripped binaries where precise type and boundary recovery is inherently more challenging, the combined context reaches a lower but still substantial 61.68\% and 72.41\%, respectively.

These results validate both the effectiveness and necessity of our context-sufficient approach. The progressive improvements support constructing minimal yet dependency-complete contexts that resolve type relations, provide precise callee prototypes, and include the required global variables. More importantly, the final leap to near-perfect compilability comes from the LLM-based repair loop, which handles residual issues (e.g., invalid casts, non-standard idioms, and imperfectly recovered structures in stripped binaries) that static context alone cannot eliminate. With this full pipeline, unstripped binaries reach 100\% compilability, while stripped binaries still achieve an impressive 99.12\% and 99.15\% success rate.

\begin{table}[t]
\renewcommand{\arraystretch}{1.2}
\caption{Ablation on compilation success rate across progressively richer contexts.
\emph{Full pipeline} includes compiler-guided LLM repair.}
\label{tab:ablation-context}
\centering
\footnotesize

\begin{tabularx}{\columnwidth}{@{}l*{6}{>{\centering\arraybackslash}X}@{}}
\toprule
\raisebox{0pt}[1.2\baselineskip][0pt]{\makecell[c]{\\ \\ \\ \\ \\ \textbf{Dataset}\\ \\ \\ \\}} &
\makecell{\;\;\textbf{PS}\,\,\,\quad\,\,\,}  &
\makecell{\;\;\textbf{LT}\,\,\,\quad\,\,\,}  &
\makecell{\textbf{LT+CP} }  &
\makecell{\textbf{LT+GV} }  &
\makecell{\textbf{LT+}\\\textbf{CP+GV} }  &
\makecell{\: \textbf{Full} \quad\quad} \\ 
\midrule
Coreutils & 0.51\% & 9.58\% & 53.44\% & 17.42\% & 69.03\% & \textbf{100\%} \\
Coreutils-stripped & 0.19\% & 12.64\% & 47.47\% & 15.43\% & 61.68\% & \textbf{99.12\%} \\
Binutils  & 1.01\% & 24.19\% & 67.80\% & 28.48\% & 80.12\% & \textbf{100\%} \\
Binutils-stripped  & 1.54\% & 21.74\% & 58.42\% & 25.55\% & 72.41\% & \textbf{99.15\%} \\
\bottomrule
\end{tabularx}

\vspace{2pt}
\raggedright\scriptsize
\textbf{Abbrev:} PS = Pseudocode only; LT = Local type definitions; CP = Callee prototypes; GV = Global variables; Full = Full pipeline (LLM repair).
\end{table}

\subsubsection{Runtime-Equivalence Ablation}
We analyze the specific contributions of our two-stage runtime feedback pipeline in achieving behavioral equivalence. Because our context reconstruction achieves high compilability, we focus on the subset of functions that compile successfully but fail initial runtime tests (\emph{Runtime Failed}). We evaluate the progressive integration of \emph{ASan-guided} repair followed by our fine-grained \emph{BP-Diff} tracing.

\textbf{Stage 1: Sanitizer-Guided Repair.} 
AddressSanitizer (ASan) provides highly accurate, localized feedback for memory safety violations. In unstripped binaries, ASan signals are relatively sparse because most decompilation errors manifest as logical discrepancies rather than memory corruptions. For instance, with GPT-5.1, ASan resolves only 20 out of 320 failed functions (6.3\%) on \textit{Coreutils} and 8 out of 153 (5.2\%) on \textit{Binutils}. 

However, on stripped binaries, the ASan repair rate noticeably increases, reaching 11.09\% (50 functions) on \textit{Coreutils-stripped} and 13.88\% (49 functions) on \textit{Binutils-stripped}. This increase highlights a specific challenge in stripped code: the imperfect recovery of data structures. Without debug symbols, fragmented or misaligned struct layouts frequently cause out-of-bounds accesses at runtime. ASan effectively catches these structural memory violations, providing the LLM with the exact bounds needed to correct the memory layout.

\textbf{Stage 2: BP-Diff Guided Repair.} 
To resolve the remaining silent semantic divergences, \sys applies Breakpoint-Matched Differential Tracing (BP-Diff). By aligning execution traces, BP-Diff supplies systematic, fine-grained evidence of divergent logic. In unstripped binaries, this deep semantic feedback drives massive improvements: the combined \emph{ASan + BP-Diff} pipeline under GPT-5.1 successfully pushes the total number of repaired functions to 245 (76.56\%) on \textit{Coreutils} and 122 (79.7\%) on \textit{Binutils}. 

Despite these strong cumulative results, achieving behavioral equivalence on stripped binaries remains fundamentally more challenging. While the full pipeline still recovers a substantial portion of failures (e.g., reaching a total of 239 repaired functions on \textit{Coreutils-stripped}, and 208 functions on \textit{Binutils-stripped} using GPT-5.1), the overall repair rate naturally drops compared to unstripped datasets. This difficulty stems from profound recovery artifacts unique to stripped code, such as misidentified parameters in indirect calls. Such structural distortions often obscure the control flow and cross-function interfaces so heavily that even fine-grained differential traces struggle to provide enough localized context for the LLM to confidently reconstruct the original semantics.

\subsection{Case Studies}

We present three representative cases spanning memory safety repair, semantic correction, and high-level structural reconstruction, which demonstrate the effectiveness of our approach, along with one challenging case that resisted automated repair.

\subsubsection{ASan-Guided Memory Violation Repair}

\begin{figure*}[!t]
  \centering
  \includegraphics[width=\textwidth]{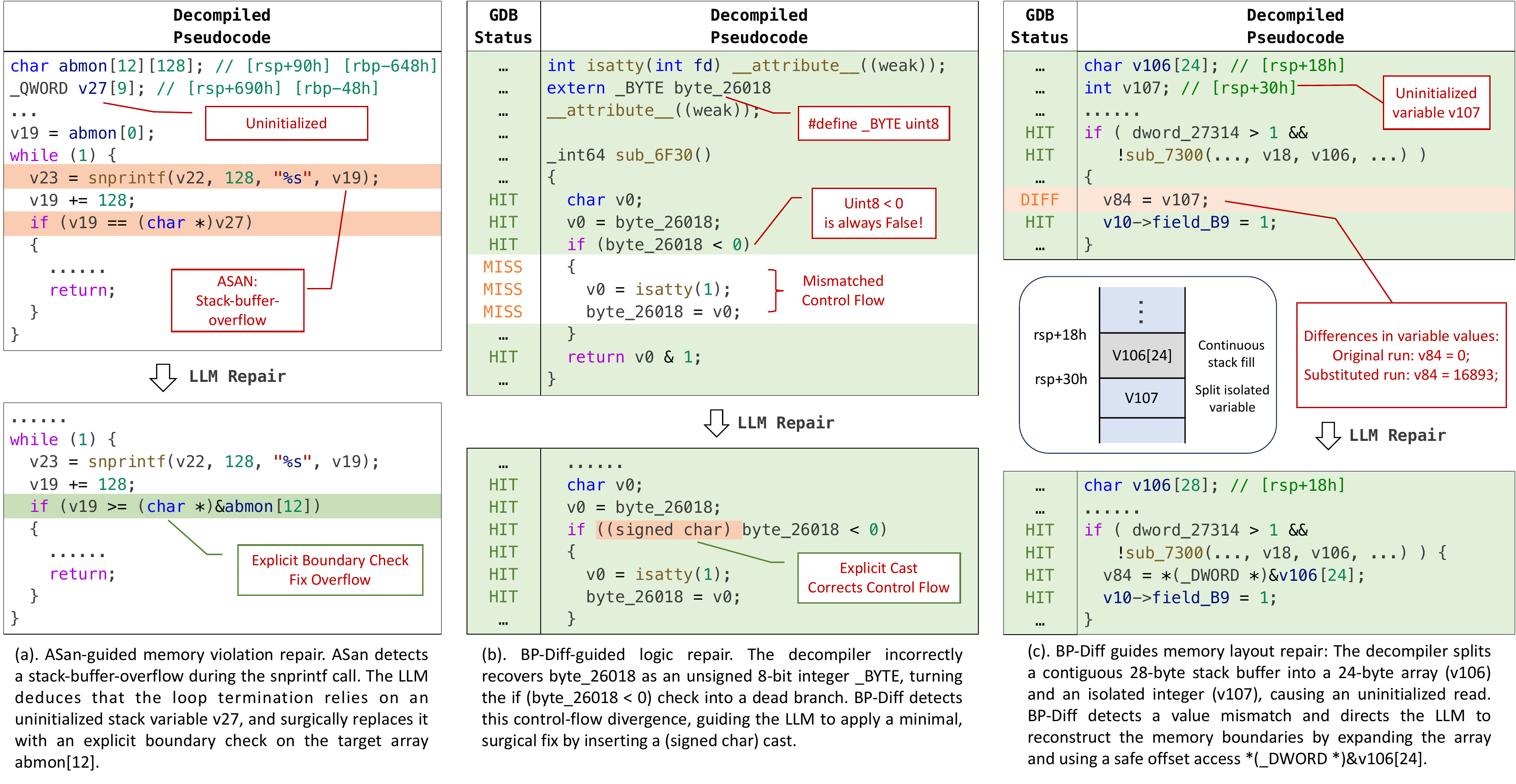}
  \caption{Case studies of runtime feedback-guided repair, including: 
  (a) \textbf{ASan-guided memory repair}, 
  (b) \textbf{BP-Diff-guided logic repair}, and 
  (c) \textbf{BP-Diff-guided memory layout repair}.}
  \label{fig:case_studies}
\end{figure*}

As shown in Figure~\ref{fig:case_studies} (a), \sys detects a stack-buffer overflow scenario during substituted execution. Specifically, AddressSanitizer flags an out-of-bounds read at the \texttt{snprintf(v22, 128, "\%s", v19)} instruction where the pointer \texttt{v19} is accessed. \sys transforms this ASan log into a structured semantic delta to guide the automated repair process. 

Upon inspecting the loop logic, \sys observes that \texttt{v19} increments by 128 (\texttt{v19 += 128}) and terminates based on a pointer comparison: \texttt{if (v19 == (char *)v27)}. Crucially, semantic analysis reveals that \texttt{v27} is completely uninitialized in the pseudocode, functioning solely as a fragile artifact of the original stack layout. Since \texttt{v19} is initially set to \texttt{abmon[0]}, the repair module infers that the loop's true intent is to iterate through the \texttt{abmon} array. 
To enforce this semantic boundary, \sys replaces the unstable stack dependency with an explicit, robust check: \texttt{v19 >= (char *)\&abmon[12]}. This surgical repair effectively eliminates the overflow and achieves behavioral equivalence without requiring large-scale code rewriting.

\subsubsection{BP-Diff--Guided Repair}
Complementing the ASan case, Figure~\ref{fig:case_studies} (b) presents a subtle logic defect that perfectly compiles and triggers no memory violations, yet fundamentally breaks the program's control flow. The decompiler incorrectly recovers a global variable \texttt{byte\_26018} as \texttt{\_BYTE} (an unsigned 8-bit integer). Consequently, the subsequent conditional check \texttt{if (byte\_26018 < 0)} always evaluates to false, turning the entire \texttt{if} block into dead code.

Since no memory corruption occurs, ASan remains silent. Instead, \sys leverages its BP-Diff tracing to align execution traces. The comparator immediately detects a critical control-flow divergence: a breakpoint inside the \texttt{if} branch is successfully hit by the original binary, but completely missed by the substituted function. 

\sys encapsulates this missing control-flow edge and the corresponding runtime state into a structured semantic delta. Guided by this precise discrepancy, the repair module easily traces the divergence back to the decompiled signedness error. To restore the correct branch evaluation, \sys applies a surgical, one-word fix by introducing an explicit type cast: \texttt{if ((signed char)byte\_26018 < 0)}. This localized edit successfully reactivates the dead branch and eliminates the behavioral divergence, proving BP-Diff's necessity for resolving silent, non-crashing semantic defects.

Beyond control-flow errors, BP-Diff effectively resolves data-flow anomalies caused by distorted memory layouts. As shown in Figure~\ref{fig:case_studies} (c), decompilers erroneously fragment contiguous buffers. Here, a 28-byte stack region is split into a 24-byte array (\texttt{v106}) and an isolated 4-byte integer (\texttt{v107}). Because the callee \texttt{sub\_7300} populates this entire region via a continuous memory fill, static analysis misinterprets \texttt{v107} as uninitialized, breaking the subsequent read.

While invisible statically, BP-Diff exposes this flaw at runtime by detecting a critical value divergence at \texttt{v84 = v107} (e.g., 16893 in the original run vs. 0 in the substituted run). Guided by this precise discrepancy and the stack context, the LLM correctly infers the boundary truncation. It applies a surgical patch that expands the array to its legitimate size (\texttt{v106[28]}) and accesses the target value via a safe offset (\texttt{*(\_DWORD *)\&v106[24]}), successfully restoring the intended continuous memory semantics.



\begin{figure}[t]
\centering
\includegraphics[width=0.85\columnwidth]{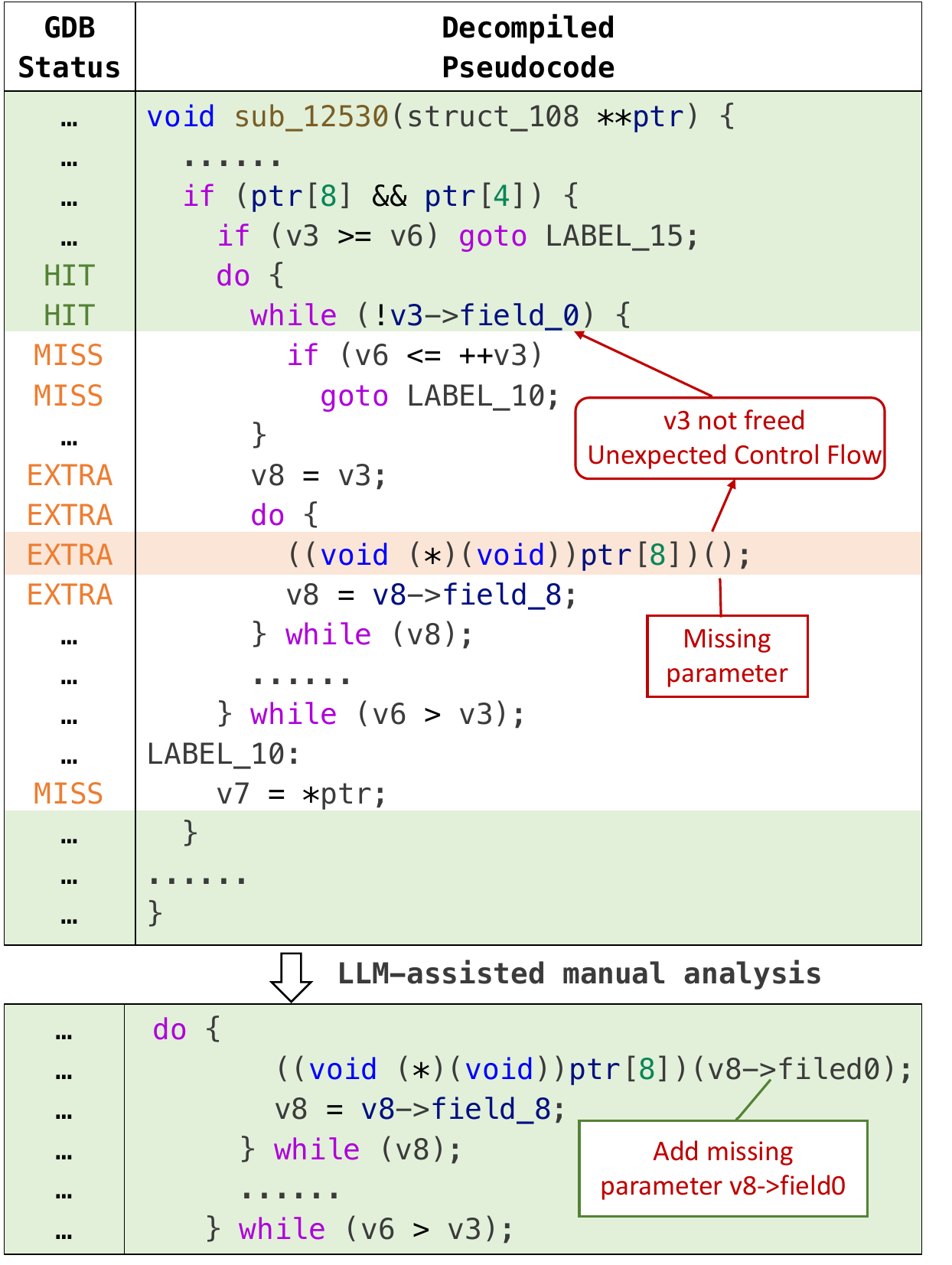}
\vspace{-2mm}
\caption{Failure Case: Opaque signature recovery in indirect calls. BP-Diff accurately detects the symptom (unexpected control flow marked by \texttt{MISS} and \texttt{EXTRA} statuses), and the LLM can suspect the indirect call as the culprit. However, recovering the exact missing parameter requires LLM-assisted manual analysis to inspect the underlying machine state.}
\label{fig:case_sig_fix}
\end{figure}

\subsubsection{Limitations and Failure Cases}
\label{subsec:limitations}
Although \sys effectively resolves many decompilation errors, stripped binaries sometimes present challenges due to the semantic gap between high-level C traces and low-level machine state. A prominent failure mode is opaque signature recovery in indirect calls. As illustrated in Figure~\ref{fig:case_sig_fix}, Hex-Rays incorrectly recovers a function pointer as taking no arguments (\texttt{((void (*)(void))ptr[8])();}), whereas the original assembly explicitly prepares an argument (\texttt{mov rdi, [rbx]}). The recompiled binary omits this setup, silently corrupting the program state.

\textbf{Value in LLM-Assisted Manual Analysis.} 
While pure C-level differential tracing struggles to fully automate this repair, BP-Diff proves highly valuable for accelerating manual root-cause analysis. As shown in Figure~\ref{fig:case_sig_fix}, \sys accurately captures the symptom: the control flow misses an expected jump (\texttt{MISS}) and falls into an unexpected execution branch (\texttt{EXTRA}). \sys leverages this trace divergence to successfully isolate the problematic indirect call. 

Although the C-level oracle abstracts away register liveness—preventing the system from autonomously deducing the exact missing parameter—this precise localization drastically narrows the search space. Guided by \sys's pinpointing, a human analyst can quickly cross-reference the original assembly, observe the \texttt{RDI} register, and manually supply the missing \texttt{v8->field\_0} argument to complete the surgical repair. 
Fully automating the resolution of such deep structural artifacts by integrating dynamic machine-state analysis (e.g., register tracking) into the repair loop remains a focus for future work.

\section{Discussion}
\label{subsec:discussion}



\textbf{Cross-Platform Adaptability.}
The current relocation engine targets Linux ELF binaries. While the core idea—binding recovered functions back into a native execution context—generalizes, adapting the system to Windows PE or macOS Mach-O demands re-engineering of relocation handling and loader integration. The conceptual framework remains applicable across platforms, but the implementation details are necessarily platform-specific.

\textbf{Signature Degradation, ABI Mismatches, and Inline Assembly.}
Static decompilers frequently fail to resolve exact function signatures and non-standard ABIs, leading to broken calling conventions. Similarly, inline assembly introduces low-level semantics that clash with standard C code. Fortunately, fine-grained runtime feedback can partially resolve these low-level semantic gaps. 

For instance, aggressive data-flow analysis can inadvertently truncate the argument list of a variadic function like \texttt{\_\_snprintf\_chk} by erroneously classifying a required argument register as dead code. The resulting recompiled code is syntactically valid but leaves the register uninitialized, causing undefined behavior during native execution. While traditional static repair cannot detect this issue, \sys leverages BP-Diff to intercept the state discrepancy surrounding the call. This dynamic evidence enables the LLM to infer the signature degradation and append the missing parameter. 

Beyond signature truncation, \sys can also resolve strict ABI mismatches when they immediately corrupt observable variables. However, more insidious violations often evade predefined breakpoint plans because the divergence manifests far from the call site. Fully resolving these elusive, deep-state bugs points to a promising future direction: empowering LLM agents with proactive, autonomous dynamic debugging capabilities to iteratively step through native execution, inspect machine state on the fly, and dynamically form hypotheses to fix complex low-level semantic gaps.

\textbf{Function Boundary Misidentification.}
While our heuristic address-to-symbol substitution effectively resolves \texttt{JUMPOUT} artifacts in standard C programs, it faces limitations with more complex or C++ binaries. In heavily optimized stripped binaries, decompilers frequently misidentify function boundaries by erroneously splitting a single function into multiple fragments. Consequently, a \texttt{JUMPOUT} instruction often targets an arbitrary mid-function address rather than a valid entry point. These fundamentally corrupted boundaries render simple symbol-table lookups and LLM-based syntax repairs futile. Fully resolving such deep structural artifacts requires advancing the state-of-the-art in binary function boundary identification prior to the decompilation phase.

\textbf{Data Layout Reconstruction and Standalone Recovery.}
While \sys achieves up to 81.48\% file-level equivalence by successfully substituting every function within a binary, this in-situ approach inherently relies on native data sections (e.g., \texttt{.data}, \texttt{.bss}). True standalone recompilation demands perfect reconstruction of the global memory layout. Decompilers often fail to capture this holistic data context, leading to severe structural losses. For example, they frequently shatter contiguous data structures into discrete global variables (e.g., \texttt{byte\_401000}, \texttt{dword\_401004}), introducing type and alignment mismatches that trigger memory corruption across function boundaries. Consequently, achieving fully independent recompilation entails repairing data, not just code. Future work will explore leveraging cross-function data-flow graphs and LLM pattern recognition to re-aggregate these fragments into high-level semantic structures, ultimately decoupling the translated source from the original binary.

\section{Conclusion}
\label{sec:conclusion}
We presented \sys, a feedback-driven decompilation framework that transforms raw pseudocode into source code that is both seamlessly compilable and behaviorally equivalent to the original binary. To overcome the compounding errors of traditional whole-binary repair, \sys isolates recompilation at the function level using globally consistent contexts. Crucially, it leverages in-situ substitutable execution to validate translated functions directly within the native binary environment. By combining AddressSanitizer diagnostics with fine-grained differential tracing, \sys establishes a robust two-stage feedback loop that systematically guides LLMs to pinpoint and resolve deep semantic divergences.

Extensive evaluations on both unstripped and fully stripped datasets of \textit{Coreutils} and \textit{Binutils} demonstrate \sys's overwhelming superiority over existing baselines. Beyond achieving up to 100\% function-level compilation success, \sys exhibits exceptional dynamic debugging capabilities. For functions that compile but fail behavioral tests, our sequential repair pipeline resolves up to 79.7\% of latent semantic bugs in unstripped binaries. Even on highly challenging stripped binaries bereft of symbolic information, \sys maintains over 99.1\% compilability and successfully repairs up to 58.9\% of initial runtime failures. This robust capability ultimately allows \sys to reconstruct the vast majority of complex real-world functions to strict behavioral equivalence.

Looking ahead, future work includes extending substitutable execution to other executable formats (e.g., PE and Mach-O) and empowering LLM agents with autonomous dynamic debugging capabilities to resolve even more complex logic defects.  Ultimately, \sys proves that fine-grained dynamic feedback effectively bridges the gap between readable pseudocode and verifiable source code, offering a viable path for the automated maintenance and modernization of binary-only software.

\ifCLASSOPTIONcompsoc
  \section*{Acknowledgments}
\else
  \section*{Acknowledgment}
\fi




\bibliographystyle{IEEEtran}
\bibliography{ref}
\appendix


\section{Loader Implementation Details}
\label{app:loader-details}

\subsection{Debugger Coordination for Relocation Engine}

\begin{figure}[h]
\centering
\includegraphics[width=0.45\textwidth]{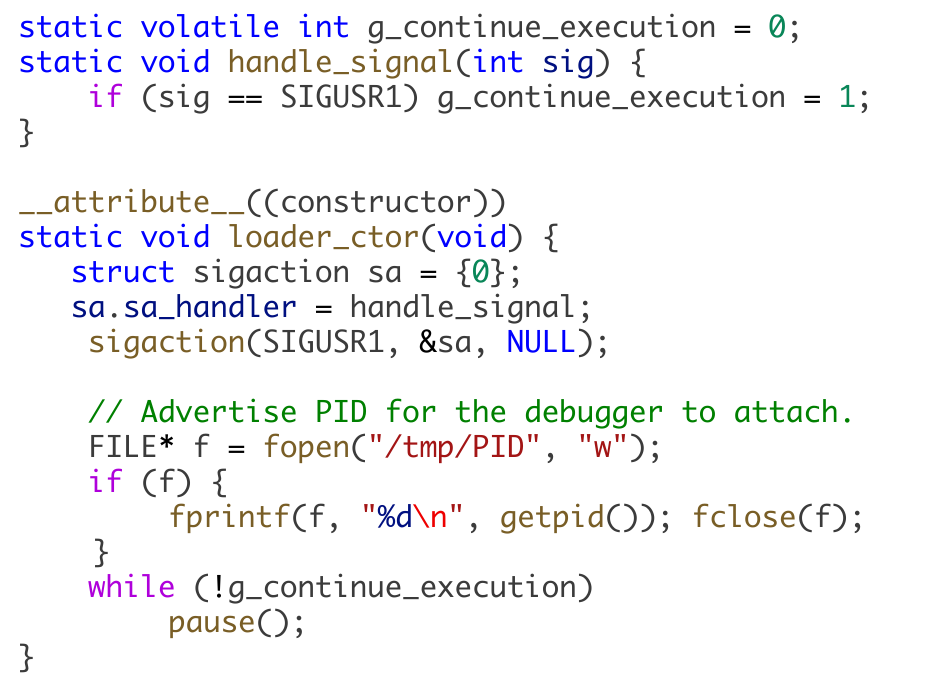}
\caption{Loader constructor that installs a signal handler, writes the process ID to a file, and pauses execution so a debugger can attach before the program proceeds.}
\label{fig:loader_debug_wait}
\end{figure}

\textbf{Startup pause and resumability.}
For debugger-assisted runs, the loader provides an optional startup barrier that allows a debugger to attach before any patching or substitution occurs. As shown in Figure~\ref{fig:loader_debug_wait}, the constructor installs a signal handler, writes the process ID to a well-known file, and then enters a pause loop. The target process resumes execution only after receiving the designated signal from the debugger, ensuring a deterministic attachment point.

\textbf{Environment switches.}
Environment variables control when this coordination logic is enabled. \texttt{BINARY\_NAME} selects the host binary in multi-process runs, and \texttt{FUNCTION\_NAME} can optionally restrict hook installation and breakpoint planning to specific symbols, reducing overhead when only a subset of functions is under study.

\subsection{Page-Protection Protocol}

\begin{figure}[h]
\centering
\includegraphics[width=0.38\textwidth]{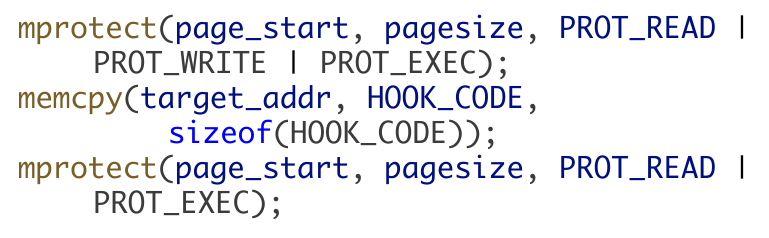}
\caption{Runtime code patching: temporarily relaxing page permissions with \texttt{mprotect}, copying hook code into the target region, and restoring execute-only protection.}
\label{fig:page_prot}
\end{figure}

To inject hooks or trampolines into existing code pages, the loader temporarily relaxes page protections, writes the patch, and then restores the original protection mode. Figure~\ref{fig:page_prot} illustrates the core sequence: a first call to \texttt{mprotect} grants read--write--execute access to the target page, the hook bytes are copied to the desired address, and a final \texttt{mprotect} restores execute-only permissions. Errors are retried with exponential backoff; persistent failures restore the prior protection before aborting the patch to avoid leaving the binary in an inconsistent state.

\section{Runtime Feedback Details}
\label{app:rt-details}

\subsection{Breakpoint and Trace Format}

\begin{figure}[h]
\centering
\includegraphics[width=0.49\textwidth]{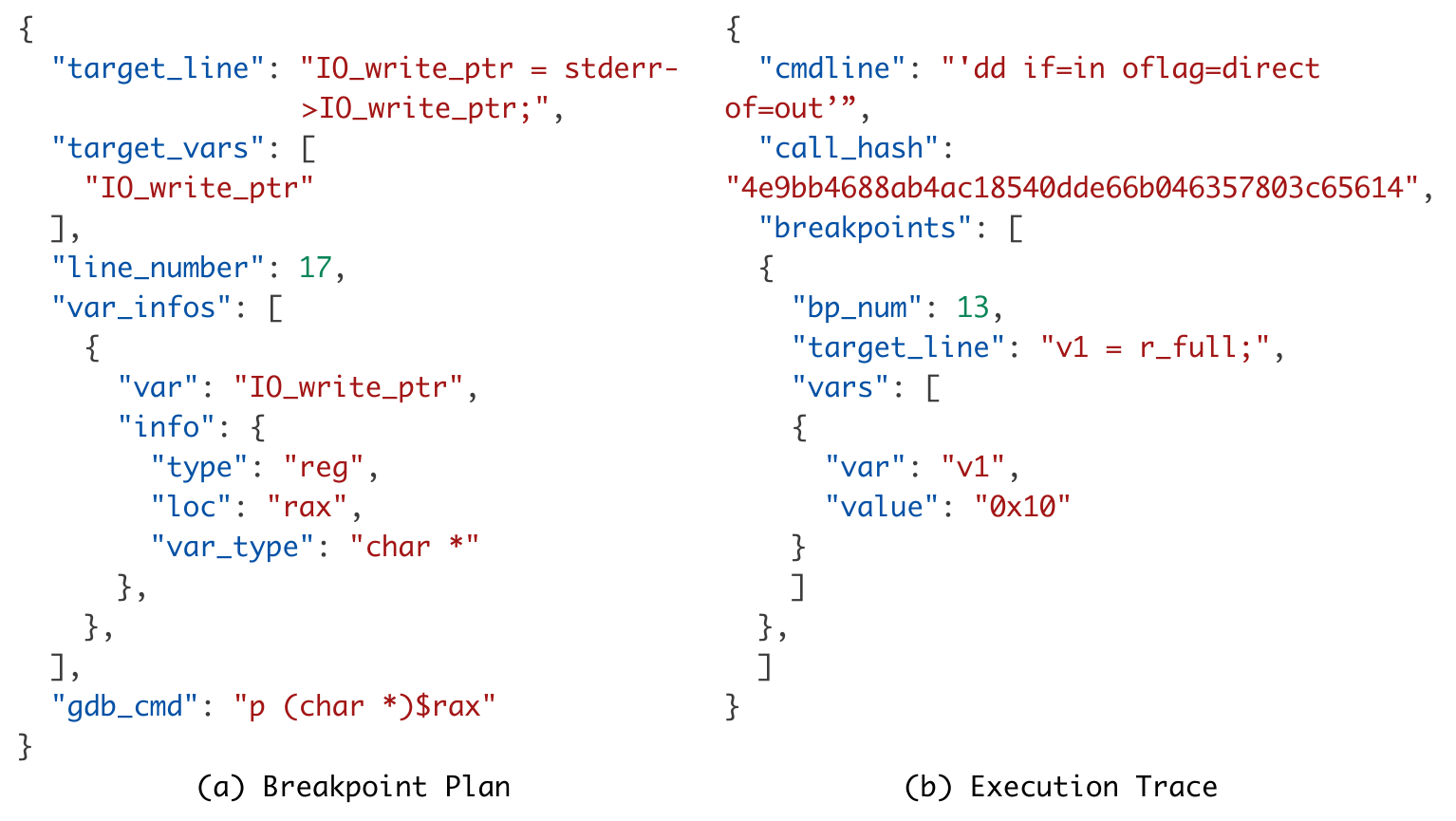}
\caption{Example of a source-level breakpoint plan (left) and a execution-trace record (right) used for breakpoint-matched differential tracing.}
\label{fig:bp_plan_trace}
\end{figure}

\textbf{Breakpoint plan.}
At the source level, each planned breakpoint site is described by a compact JSON record that includes the target source line, the variables to observe, and how to obtain their values in GDB (e.g., via register names or local-variable slots). An example is shown on the left of Figure~\ref{fig:bp_plan_trace}.

\textbf{Execution trace records.}
During in-situ execution, each breakpoint hit produces a corresponding JSON fragment that records the command line (or test identifier), a call hash, the breakpoint number and the observed values of watched variables. The right side of Figure~\ref{fig:bp_plan_trace} shows a representative trace entry.

\subsection{Normalization and Delta Extraction}

\textbf{Noise suppression.}
To avoid spurious differences, traces are normalized before comparison: pointer identities, ASLR-dependent addresses, randomized iteration orders, and timestamps are masked or canonicalized.

\textbf{Pairing.}
Original and substituted traces are paired by input hash and call depth, and within each pair, breakpoint hits are aligned by breakpoint identifier and invocation index. The alignment tolerates benign reordering of commutative outputs (e.g., independent debug prints).

\textbf{Delta encoding.}
From the aligned traces, \sys extracts a structured delta that summarizes behavioral differences at each site, including missing or extra breakpoint hits, mismatched variable values, and divergent control-flow progress. These deltas, together with the surrounding source context, are passed to the LLM as fine-grained guidance for subsequent repair iterations.

\subsection{LLM-Based Repair Executor}
\label{subsec:repair_executor}

While the high-level methodologies for compiler-guided correction and differential tracing repair are detailed in the main text, the following sections provide the exact context, instructions, and templates provided to the LLM to ensure effective, high-quality code patches.

\textbf{Compiler-Guided Correction Prompt.} We parse compiler diagnostics to extract error classifications and line-specific spans. The resulting prompt provides the agent with line-numbered source snippets, localized around reported errors in large functions to maintain context stability. As specified in the following instruction block, the agent is tasked with proposing minimal necessary changes while preserving public interfaces. To address recurring portability issues, the prompt provides guidance on common challenges such as SIMD intrinsics and assembly idioms:

\begin{promptbox}[title=Patch generation and application]
You are an expert C++ code compiler, skilled at resolving compilation errors. Your task is to correct the given decompiled pseudo code, to eliminate G++ compilation errors.

\textbf{Key context:}
\begin{enumerate}
    \item This is a C++ file, containing function implementations that have been decompiled from binary code.
    \item You must resolve all errors by identifying and precisely locating each issue mentioned in the compiler output.
    \item For each identified issue, determine which file(s) contain the error (e.g., \textless file\_path1\textgreater\ or \textless file\_path2\textgreater).
    \item Do not add header files.
    \item You can follow these examples when fixing compilation errors: 
\end{enumerate}

\textbf{a.} Only modify what is necessary and avoid introducing any undeclared variables. If you change a variable, ensure the change is reflected in all subsequent uses.

\textbf{b.} Replace SIMD intrinsic instructions with standard memory copy operations. IDA often decompiles optimized memory copies as SSE/AVX instructions.

\textbf{c.} When you encounter an error like this: \textit{invalid conversion from 'uint64*' \{aka 'long long unsigned int*'\}} to 'uint64. You can add a forced type conversion.

\textbf{d.} When you encounter redefinition errors for global variables (e.g. \textit{redefinition of ‘idx\_t nslots’}), simply remove the duplicate declarations.

\textbf{e.} If the compiler reports ambiguating new declaration and shows a system prototype (e.g. \texttt{void *reallocarray(...)}), use the system header's prototype or remove the redundant local declaration.
\end{promptbox}


\textbf{Differential Tracing Repair Prompt.} This prompt is designed to guide AI agents in analyzing the dynamic execution trace divergences between the original function (\textbf{Binary A}: the original baseline reference) and the substituted function (\textbf{Binary B}: the version reconstructed from decompiled sources) to locate and repair logic or data structure discrepancies within the code.

\begin{promptbox}[title=Patch generation and application]
You are an expert C/C++ reverse engineer and software developer, familiar with HexRays decompiler output and GDB debugging. \\
Your goal is to align the behavior of the function in binary b with that in binary a, assuming the down/up behaviors are already aligned.

\textbf{Setup recap:}
\begin{itemize}
    \item Binary a is the original executable; binary b is rebuilt from decompiled sources.
    \item Breakpoint comparisons are organized into blocks by type:
    \begin{itemize}
        \item \texttt{[Block: All Same]}: All breakpoints match with same variable values (shows first 2 and last 2)
        \item \texttt{[Block: Diff Vars]}: Breakpoints match but some variable values differ (shows each diff\_var with context)
        \item \texttt{[Block: Only Binary A]}: Breakpoints hit only in binary\_a, missing in binary\_b (shows first 2 and last 2)
        \item \texttt{[Block: Only Binary B]}: Breakpoints hit only in binary\_b, extra execution (shows first 2 and last 2)
        \item \texttt{[SIGNAL ]}: A line indicating a signal or crash point, need to pay special attention sometimes.
    \end{itemize}
    \item Control-flow divergences (Only Binary A / Only Binary B) are often symptoms of earlier data corruption. You \textbf{MUST} always inspect the \texttt{[Block: Diff Vars]} immediately preceding a control-flow split. Do not alter branch conditions or insert early returns unless you are absolutely sure the data flow is correct.
\end{itemize}

\textbf{Failures summary:} \\

Divergence windows (ordered by first divergence per call): \\
Call 1 @ 'du -s gt-long-from-unreadable.sh/'

\begin{verbatim}
[Block: Diff Vars] (16 breakpoints)
  [=] L49: v18 = v2;
  [!] L50: v14 = (char *)v5 + 16*v4; 
  (v14: a=0x80, b="")
  [=] L51: v6 = *(_QWORD *)&a1->pad_20[16];
  ... (3 same_var breakpoints omitted)
  [=] L57: v21 = *(_QWORD *)&a1->pad_20[32];
  [!] L58: v22 = a1->f_48; 
  (a={0,NULL}, b={<NULL>,<NULL>})
  [SIGNAL] SIGSEGV L59: v8 = sub_BA60(
  (struct_2626 *)&ptr, a1, 0);
    Backtrace (condensed):
    #0  0x000000000000003b in ?? ()
    #2  0x... in sub_C390 (a1=0x..., a2=0x2e) 
    at sub_C390.cpp:59
    ... (omitting libc frames)

[Block: Only Binary A] (7 breakpoints)
  [!] L59: v8 = sub_BA60((struct_2626 *)&ptr, 
  a1, 0); (v8=1)
  [!] L62: free(a1->field_0);
  ...
  [!] L66: a1->field_18 = v16; 
  (a1->field_18=0x18)
\end{verbatim}
Source excerpt from \texttt{sub\_C390.cpp} with line numbers (trimmed around the divergence windows): \\
\texttt{[All 84 lines]} \\
...... \\
\texttt{13$\rightarrow$\_\_int64 sub\_C390(struct\_2626 *a1, struct\_2626 *a2)} \\
\texttt{14$\rightarrow$\{ } \\
\texttt{15$\rightarrow$ struct\_2629 *v2; // r12} \\
......

Relevant struct definitions from header:
\begin{verbatim}
struct struct_2629
{
    int field_0;
    int field_4;
    int field_8;
    unsigned __int8 pad_C[4];
    char field_10;
};
\end{verbatim}
......

\subsection*{Diagnostic Focus}
Before editing, \textbf{prioritize identifying root causes} of divergence. \textbf{DO NOT simply force binary\_b to match binary\_a's behavior—find and fix the HexRays decompiler root cause.}
\end{promptbox}

\textbf{Patch generation and application.}
Each patch specifies a target file path, a \texttt{search} block, and a \texttt{replace} block.
If the \texttt{search} block matches in the target file, the matched region is replaced verbatim by the \texttt{replace} block.
All edits must respect the constraints stated above (minimal change, interface stability, no new external dependencies).

\begin{promptbox}[title=Patch generation and application]
Response requirements:
\begin{enumerate}
    \item In \textbf{Analysis}, pinpoint the earliest real divergence and explain why it happens.
    \item In \textbf{Fix}, output the \textbf{necessary search/replace} patch for all diff windows. \textbf{You MUST produce at least one SEARCH/REPLACE block.}
\end{enumerate}

\texttt{cpp\_path}

\begin{verbatim}
<<<<<<< SEARCH
(original code fragment)
=======
(corrected code fragment)
>>>>>>> REPLACE
\end{verbatim}

\begin{enumerate}
    \setcounter{enumi}{2}
    \item In \textbf{Explanation}, justify how the patch resolves the divergence without breaking other logic.
\end{enumerate}

Fixing rules:\\
\# \textit{SEARCH/REPLACE block} Rules:

Every \textit{SEARCH/REPLACE block} must use this format:
\begin{enumerate}
    \item The \textit{FULL} file path alone on a line, verbatim. No bold asterisks, no quotes around it, no escaping of characters, etc.
    \item The opening fence and code language, eg: \texttt{```cpp}
    \item The start of search block: \texttt{<<<<<<< SEARCH}
    \item A contiguous chunk of lines to search for in the existing source code
    \item The dividing line: \texttt{=======}
    \item The lines to replace into the source code
    \item The end of the replace block: \texttt{>>>>>>> REPLACE}
    \item The closing fence: \texttt{```}
\end{enumerate}

Every \textit{SEARCH} section must \textit{EXACTLY MATCH} the existing file content, character for character, including all comments, docstrings, etc.\\

\textit{SEARCH/REPLACE} blocks will \textit{only} replace the first match occurrence.\\
Including multiple unique \textit{SEARCH/REPLACE} blocks if needed.\\

Keep \textit{SEARCH/REPLACE} blocks \textbf{concise}.\\
Break large \textit{SEARCH/REPLACE} blocks into a series of smaller blocks that each change a small portion of the file.\\
Include just the changing lines, and a few surrounding lines if needed for uniqueness.\\

Only create \textit{SEARCH/REPLACE} blocks for files that the user has added to the chat!

To move code within a file, use 2 \textit{SEARCH/REPLACE} blocks: 1 to delete it from its current location, 1 to insert it in the new location.

\end{promptbox}





%



\end{document}